\documentclass[12pt]{elsarticle}

\usepackage{amssymb}
\usepackage{amsmath} 
\usepackage{fullpage} 
\usepackage{multicol}
\usepackage{breqn}
\usepackage{graphicx}
\usepackage{color}

\biboptions{numbers,sort&compress}

\begin{document}

\begin{frontmatter}

\title{Bayesian Uncertainty Quantification and Information Fusion in CALPHAD-based Thermodynamic Modeling}

\author{P. Honarmandi$^a$\corref{cor1}} 
\ead{hona107@tamu.edu}
\author{T. C. Duong$^{a,b}$\corref{cor2}}
\author{S. F. Ghoreishi$^b$\corref{cor2}}
\author{D. Allaire$^b$\corref{cor2}}
\author{R. Arroyave$^{a,b}$\corref{cor2}}

\cortext[cor1]{Corresponding author at: Texas A\&M University, Department of Materials
Science and Engineering, 3003 TAMU, College Station, TX 77843-3003, USA.}

\address{$^a$Department of Materials Science and Engineering, Texas A\&M University, College Station, Texas, USA}

\address{$^b$Department of Mechanical Engineering, Texas A\&M University, College Station, Texas, USA}

\begin{abstract}

Calculation of phase diagrams is one of the fundamental tools in alloy design---more specifically under the framework of Integrated Computational Materials Engineering. Uncertainty quantification of phase diagrams is the first step required to provide confidence for decision making in property- or performance-based design. As a manner of illustration, a thorough probabilistic assessment of the CALPHAD model parameters is performed against the available data for a Hf-Si binary case study using a Markov Chain Monte Carlo sampling approach. The plausible optimum values and uncertainties of the parameters are thus obtained, which can be propagated to the resulting phase diagram. Using the parameter values obtained from deterministic optimization in a computational thermodynamic assessment tool (in this case Thermo-Calc) as the prior information for the parameter values and ranges in the sampling process is often necessary to achieve a reasonable cost for uncertainty quantification. This brings up the problem of finding an appropriate CALPHAD model with high-level of confidence which is a very hard and costly task that requires considerable expert skill. A Bayesian hypothesis testing based on Bayes' factors is proposed to fulfill the need of model selection in this case, which is applied to compare four recommended models for the Hf-Si system. However, it is demonstrated that information fusion approaches, i.e., Bayesian model averaging and an error correlation-based model fusion, can be used to combine the useful information existing in all the given models rather than just using the best selected model, which may lack some information about the system being modelled. 

\end{abstract}


\end{frontmatter}

\begin{multicols}{2}

\section{Introduction}
\label{1}
 
Uncertainty quantification (UQ) and its propagation (UP) across multi-scale model/experiment chains are key elements of decision-based materials design in the framework of Integrated Computational Materials Engineering (ICME), where databases, multi-scale modeling and simulation tools, and experiments are integrated with the aim of time reduction in design and manufacturing of materials/products~\cite{council_2008}. In this context, the understanding and quantification of uncertainties can provide a confidence measure for the applicability of models for decision making in materials design. 

Generally, UQ incorporates the detection of uncertainty sources and the development of corresponding appropriate mathematical approaches to calculate the error bounds for any quantity of interest in models ~\cite{panchal_2013,chernatynskiy_2013,ghoreishi2017adaptive}. The uncertainty can arise from different sources, which are categorized as natural uncertainty (NU) due to the random nature of a physical system, model parameter uncertainty (MPU) resulting from the lack of sufficient and/or accurate data for model parameters, propagated uncertainty (PU) in the case of multi-scale modeling, and model structure uncertainty (MSU) owing to any simplifications, assumptions, and/or incomplete physics in the model~\cite{choi_2008}. UP also refers to the determination of model output uncertainties based on the uncertainties of its input variables for either individual or multi-scale models~\cite{panchal_2013, ghoreishi2016uncertainty,ghoreishi2016compositional}. 

Despite the importance of UQ/UP, there are comparably few works applied to problems in (computational) materials science. Chernatynskiy et al.\ \cite{chernatynskiy_2013} have reviewed these efforts, focusing on UQ problems related to atomic-scale as well as multi-scale simulations. For over a decade, at the electronic/atomic scale,  probabilistic parameterization in first principles calculations based on density function theory (DFT) \cite{mortensen_2005,hanke__2011} remains as some of the most significant works on UQ applied to materials problems. Aldegunde et al. \cite{aldegunde_2016} have recently used a machine learning-informed Bayesian approach
to quantify uncertainties associated to the prediction, via cluster expansions, of the thermodynamic properties of alloys. In that work, the resulting uncertainties are associated to the model parameters as well as the structure of the models themselves, which results from poorly converged cluster expansions due to lack of training data. In another work, the diffusivity error in molecular dynamic simulations of a Ni/Al nano-laminate bilayer has been assessed using a Markov Chain Monte Carlo (MCMC) method in the context of Bayesian statistics~\cite{rizzi_2011}. In regard to UQ in multi-scale simulations, Volker et al.~\cite{volker_2011} have used DFT results for a phase field model to provide a connection between atomic- and meso-scale descriptions of ferro-electric materials, and also applied a sensitivity analysis to identify the most significant parameters of the free energy functional and to quantify the uncertainties in the simulations. Moreover, Liu et al.~\cite{liu_2009} have utilized a Bayesian stochastic approach for probabilistic parameter calibration and a stochastic projection with polynomial chaos expansions for the propagation of their uncertainties across multi-scale constitutive models that link the microstructure, property, and performance of random heterogeneous composite materials. It is worth mentioning that there is also some UQ work for multi-scale modeling of plasticity (deformation) in poly-crystalline materials~\cite{kouchmeshky_2010,koslowski_2011,salehghaffari_2012}.

While there are emerging several efforts centered on UQ/UP applied to materials simulations, UQ applied to CALculation of PHAse Diagrams (CALPHAD) methodologies remains poorly explored. This is unfortunate as CALPHAD-based thermodynamic descriptions are the foundation of most proposed ICME/alloy design frameworks. UQ in the calculation of equilibrium of phase diagrams is crucial since any small variations in chemical composition and/or temperature due to their uncertainties can alter the (predicted) stability, volume fraction, and chemical composition of microstructural phases and may affect the materials properties considerably. For example, if the stability of a phase should be suppressed in the design to prohibit its detrimental effects on the final product, the design space for this particular materials system should lie outside the boundaries representing the onset of stability of such detrimental phases but also out of their uncertainty bounds in order to have more confidence in the design. 

Focused work on UQ applied to CALPHAD-based thermodynamic assessments remains scarce~\cite{konigsberger_1991,olbricht_1994,chatterjee_1994,malakhov_1997,chatterjee_1998,acker_1999}. However, Stan and Reardon~\cite{stan_2003} and Otis and Liu~\cite{otis_2017} have recently proposed approaches that constitute increasingly sophisticated approaches to UQ applied to CALPHAD. In the former study~\cite{stan_2003}, a fuzzy logic weighted genetic algorithm (GA) was applied as a sampling tool in a Bayesian-based framework to find the free energy parameters' posterior probability distribution given some uncertain thermodynamic data, and then the parameters' uncertainties were propagated to the resulting phase diagram by sampling from the posterior probability distribution. In the later work~\cite{otis_2017}, an ICME-directed multi-scale (or rather multi-level) modeling has been introduced that links high-throughput first-principles calculations to CALPHAD modeling. The most relevant parameters in the sublattice-based CALPHAD modeling of that work were determined using both Akaike Information Criterion (AIC) and F-test, and a Markov chain Monte Carlo (MCMC) sampler was used to quantify the parameters' posterior probability distribution.

Furthermore, Duong et al.~\cite{duong_2017} have analyzed the uncertainties in the case of a multi-scale modeling for pseudo-binary Ti$_{2}$AlC-Cr$_{2}$AlC MAX phases, which links first principles calculations and the CALPHAD method. In that study, the overall uncertainties have been reflected on the calibrated thermodynamic parameters through an MCMC approach, and then propagated to the quasi-binary phase diagram. Earlier, Duong et al.~\cite{duong_2016} employed UQ to investigate the effect of uncertainties in the phase stability of the U-Nb system. However, a detailed and thorough analysis of uncertainty and selection of models with appropriate parameters is still lacking in CALPHAD modeling.

In this work, a thorough evaluation of the uncertainty is performed for not only CALPHAD model parameters, but also for their resulting phase stability diagrams using the Hf-Si binary system as a case study. It should be noted that calculation of Hf-Si binary phase diagram and its uncertainties is of great importance since adding Hafnium to Niobium silicide based alloys (as promising turbine airfoil materials with high operating temperature) increases their strength, fracture toughness, and oxidation resistance significantly \cite{zhao_2000,yang_2003,zhao_2001}. In this regard, an MCMC-Metropolis Hastings algorithm and a forward analysis of parameters' posterior samples are applied for the quantification of the parameters' uncertainties and their propagation to resulting phase diagrams, respectively. In this paper, it is also shown how to select the most relevant CALPHAD model given experimental data for a system based on Bayesian hypothesis testing. For this analysis, four expert-proposed CALPHAD models for the Hf-Si system are considered. 

Since finding the appropriate number of parameters in CALPHAD models---or rather, the appropriate model parameter set---is a hard and skillful task, it is necessary to propose systematic approaches that can provide sufficiently reliable results with no need for the manual identification of the best predicting model. In this regard, randomly selected CALPHAD models can be combined intelligently together through information fusion approaches, instead of being wasted in search for the best model. In our work, Bayesian model averaging (BMA) and an error correlation-based model fusion (CMF) approach are used to combine all the given model results together in different ways with their own specific purposes. In BMA, each model has some probability of being true and the fused estimate is a weighted average of the models. This method is extremely useful in the case of model-building process based on a weighted average over the models' responses, and/or less risk (more confidence) in design based on broader uncertainty bands provided by a weighted average over the uncertainties of the models' responses. On the other hand, the information fusion technique based on the correlations between the model deviations is applied for the purpose of acquiring more precise estimations and lower uncertainties compared to results obtained from each applied individual model.
 
\section{CALPHAD Model Description}
\label{2}
 
Four sets of models describing Gibbs free energies of the binary system are considered in the current work. In each of these sets, the intermetallic compounds are described using the line-compound formalism \cite{hillert_2001}, which reads:

\begin{dmath}
\label{eq 1}
G^{Hf_k Si_l}=\frac{k}{k+l} {^0}G_{Hf}^{HCP}+\frac{l}{k+l} {^0}G_{Si}^{Diamond}+a+bT,
\end{dmath}
where $k$ and $l$ are the compound numbers, ${^0}G_{Hf}^{HCP}$ and ${^0}G_{Si}^{Diamond}$ are the chosen energy references corresponding to the energies of pure HCP-Hf and diamond-Si as given in the SGTE database~\cite{dinsdale_1991}, $a$ and $b$ are model parameters, and $T$ is temperature (in Kelvin).

 The liquid phase is described using the sub-regular solid solution model as follows:
 
\begin{dmath}
\label{eq 2}
G^{Liq} = \sum\limits_{i=1}^{N} x_{i}{^{0}G^{Liq}_{i}} + RT\sum\limits_{i=1}^{N} x_{i}\ln{x_{i}} + \sum\limits_{i=1}^{N} x_{i} \sum\limits_{j\neq i}^{N} x_{j} \sum\limits_{n=0}^{M} {^n}L_{ij}(x_i-x_j)^n,
\end{dmath} 
where $x_i$ is the mole fraction of the constituent $i$ (either Si or Hf), ${^0}G^{liq}_{i}$ is the constituent energy taken, again, from the SGTE database~\cite{dinsdale_1991}, $T$ is temperature (in Kelvin), $R$ is the gas constant, and ${^n}L_{ij}$ is given as: 

\begin{dmath}
\label{eq 3}
{^\nu}L_{ij} = {^\nu}a_{ij} + {^\nu}b_{ij}T,
\end{dmath}
where ${^\nu}a_{ij}$ and ${^\nu}b_{ij}$ are model parameters, which describe the interactions between the constituents beyond those of ideal mixing.

The terminal phases, namely HCP-Hf (A3) and diamond-Si (A5), are described by either the line-compound formalism \cite{hillert_2001} or the sub-regular solid solution model.

In the current work, four models with 17, 20, 28, and 30 parameters identified as models 1 through 4 are chosen for Hf-Si binary system by expert opinion to show how the model selection and model fusion approaches work in this case study. As mentioned earlier, these parameters enter the polynomial coefficients resulting from various orders of polynomial expansion in the description of Gibbs free energy for the liquid and the end members of the phases (dilute phases that are close to pure metals) in the system, i.e., $a$s and $b$s in Equation~\ref{eq 3}. The orders of the (Redlich-Kister)  polynomial expansion associated with the liquid and the end members are listed in Table~\ref{Table 1} for all the given models. According to this table, it can be observed that the parameters in the smaller models are subsets of the parameters in the larger models. 

\begin{table*}[htp]
\centering
\caption{Orders of polynomial expansion ($n$) in Gibbs energies of the liquid and the end members for each given model}
\setlength{\arrayrulewidth}{0.7mm}
\setlength{\tabcolsep}{15pt}
\renewcommand{\arraystretch}{1.2}
{\begin{tabular}{ c c c c }
\hline\hline
{} & Liquid & $(HCP-Hf)$/$(BCC-Hf)$/$(Si)$ & Intermetallic Phases \\
\hline
Model 1 & 2 & 1 & 2 \\
Model 2 & 2 & 2 & 2 \\
Model 3 & 4 & 4 & 2 \\
Model 4 & 6 & 4 & 2 \\
\hline\hline
\end{tabular}}
\label{Table 1}
\end{table*}

\section{Uncertainty Quantification Methodology}
\label{3}

\subsection{Markov Chain Monte Carlo-Metropolis Hastings Algorithm}
\label{3-1}

In our work, the MCMC Metropolis Hastings toolbox in Matlab has been utilized for probabilistic calibration of the parameters in the applied CALPHAD models. After the introduction of prior knowledge for parameters (initial values (vector $\theta^0$), lower and upper bounds, and probability density functions (PDFs) in this algorithm, parameter vectors are randomly sampled from a non-stationary proposal posterior probability distribution, which is an arbitrary multivariate Gaussian distribution with a mean value at $\theta^0$ for the first sampling or the last accepted parameter vector during the next parameter sampling. The sampled parameter vector in each iteration can be accepted or rejected through a criterion in the context of the Bayesian statistic, known as the Metropolis Hastings (M-H) ratio:

\begin{dmath}
\label{eq 4}
\textrm{M-H}=\frac{p(\theta^{cand})p(D|\theta^{cand})}{p(\theta^i)p(D|\theta^i)} \frac{q(\theta^i|\theta^{cand})}{q(\theta^{cand}|\theta^i)},
\end{dmath}
where $\theta^{i}$, $\theta^{cand}$, and $D$ are the last accepted sample of parameter vector, the new sample of parameter vector as a candidate, and the given data, respectively. 

Based on Bayes' theorem, the joint (posterior) probability in each case is proportional to the prior probability times the likelihood, $p(\theta)p(D|\theta)$. The likelihood is the conditional probability of the data, $D$, given the parameter vector, $\theta$, and in this work is considered as a Gaussian distribution centered at the data, $D$, with a variance, $\sigma^2$, determined by the data error. This variance can be updated as a hyper-parameter during MCMC sampling when the data error is unknown. In this case, the variance samples are generated from an inverse gamma posterior PDF resulting from the introduction of a non-informative inverse gamma prior PDF for the variance $\sigma^2$~\cite{duong_2016,gelman_2014}. The second ratio in Equation~\ref{eq 4} is the Hastings ratio, which considers the asymmetric effect of the proposal probability distribution in the acceptance/rejection criterion of the parameter vector.

After the calculation of the M-H ratio for each sampling iteration, if $\min\{\textrm{M-H},1\} \times 100$ is greater than 1, the candidate for the parameter vector is accepted as the new sample; otherwise, the candidate may still be accepted with a probability of $\min\{\textrm{M-H},1\} \times 100$ \cite{billera_2001}. In the case of the candidate rejection, the last accepted parameter vector is repeated in the MCMC chain as the new sample. At the end of the MCMC sampling process, a chain of the parameter vectors is obtained as $\big\{\theta^0,...,\theta^N\big\}$ whose mean values and variance-covariance matrix after removal of "burn-in period" demonstrate the plausible optimum values of the parameters and their overall reflected uncertainties, respectively. It should be noted that the parameters' convergence during MCMC sampling algorithm is defined as the convergence of the parameters' cumulative mean values towards almost constant values, which is equivalent to the fluctuation of the parameters' samples around their mean values across MCMC sampling \cite{lynch_2007}.

\subsection{Bayesian Model Selection/Averaging}
\label{3-2}

Jeffreys \cite{jeffreys_1935} proposed a hypothesis testing method in the context of Bayesian statistics that evaluates the evidence in favor of a scientific theory or hypothesis. Models are considered as hypotheses in the case of Bayesian model selection (BMS). In BMS, non-nested models (models with at least one uncommon parameter) can also be compared together, which is very difficult or sometimes impossible through frequentist approaches~\cite{kass_1995}. Generally, models/hypotheses with greater posteriors are more favored by the evidence and considered as better models/hypotheses. Accordingly, the Bayesian comparison criterion for models/hypotheses is defined as the posterior odds of each two applied models/hypotheses given data. For instance, posterior odds of model one ($M_1$) to model two ($M_2$) given the data ($D$) is expressed as follows:
\begin{dmath}
\label{eq 5}
\frac{p(M_1|D)}{p(M_2|D)}=\frac{p(D|M_1)}{p(D|M_2)} \frac{p(M_1)}{p(M_2)}=B_{12}\lambda_{12}.
\end{dmath}
Bayes' factor is the ratio of the marginal likelihoods and usually suggests which model/hypothesis is more favored by the evidence (data):
\begin{equation}
\label{eq 6}
p(D|M_k)=\int p(D|\theta_k, M_k)p(\theta_k|M_k) d\theta_k,
\end{equation}
where $\theta_k$ is the parameter vector under $M_k$. Therefore, the key element in BMS is the calculation of the marginal likelihoods offered by Equation~\ref{eq 6}, which is challenging in the case of high dimensional $\theta_k$. There are different methods to approximate these integrals, which include Laplace's method, the Schwarz criterion, simple Monte Carlo, importance sampling, adaptive Gaussian quadrature, and simulating approaches from the parameters' posterior \cite{kass_1995}. In our work, the MCMC sampling method is used to simulate from the parameters' posterior for the integral estimation.

In the Monte Carlo method, the integral in Equation~\ref{eq 6} is approximated as:

\begin{dmath}
\label{eq 7}
p(D|M_k)\approx \frac{1}{N}\sum_{i=1}^{N} p(D|\theta^{(i)}_k, M_k),
\end{dmath}
where $\theta^{(i)}_k$s are samples from the parameters' prior $p(\theta_k|M_k)$. This approximation is equivalent to the average of the likelihood values obtained from the sampled parameters. However, this approach can be fairly inefficient in the case of a more concentrated posterior distribution compared to the prior distribution since the likelihood values for most of the sampled $\theta^{(i)}_k$ will be very small and the approximation result will be considerably weighted by a few samples with large likelihood values. For this reason, importance sampling techniques are usually applied \cite{kass_1995}:

\begin{dmath}
\label{eq 8}
p(D|M_k) \approx \frac{1}{N}\sum_{i=1}^{N} \frac{p(D|\theta^{(i)}_k, M_k)p(\theta^{(i)}_k|M_k)}{p^*(\theta^{(i)}_k)},
\end{dmath}

\begin{dmath}
\label{eq 9}
p(D|M_k) \approx \frac{1}{N}\sum_{i=1}^{N} \frac{p(D|\theta^{(i)}_k, M_k)w^{(i)}_k}{w^{(i)}_k},
\end{dmath}
where $p^*(\theta_k)$ is a suitable distribution that enables the sampling of the important parameter values for the sake of more efficient sampling. In Equation \ref{eq 9}, $w^{(i)}_k= \frac{p(\theta^{(i)}_k|M_k)}{p^*(\theta^{(i)}_k)}$, which can provide a weighted average of likelihood values obtained using the sampled parameters from $p^*(\theta_k)$.

In the case that the importance sampling function is proportional to the posterior probability density function of the parameters in model $k$, i.e., $p^*(\theta_k)=p(D|\theta_k, M_k)p(\theta_k|M_k)$, Equation~\ref{eq 9} is turned into:

\begin{dmath}
\label{eq 10}
p(D|M_k) \approx \Bigg\{\frac{1}{N}\sum_{i=1}^{N} p(D|\theta^{(i)}_k, M_k)^{-1}\Bigg\}^{-1}
\end{dmath}
MCMC is usually used to sample $\theta^{(i)}_k$s from ($p^*(\theta_k)$). Equation \ref{eq 10} shows a harmonic mean of likelihood values obtained from the MCMC sampled parameters after the removal of the burn-in period.

In addition to model selection, Bayes' factors can be used to determine the posterior probability distribution of the competing models given data, which can be defined as their associated weight in the context of Bayesian model averaging (BMA). Generally, BMA can provide a combined inference from all the competing models that can be very useful for model-building process or less risky predictions in design. In this approach, the posterior density (weight) associated with each model can be obtained as follows:

\begin{dmath}
\label{eq 11}
p(M_k|D)=\frac{p(D|M_k)p(M_k)}{\sum_{i=1}^{K} p(D|M_k)p(M_k)} \\
=\frac{B_{k1}\lambda_{k1}}{\sum_{i=1}^{K} B_{i1}\lambda_{i1}}=\frac{B_{k1}}{\sum_{i=1}^{K} B_{i1}},
\end{dmath}
where $K$ is the total number of models. In this equation, $M_1$ is considered as reference for the calculation of all the Bayes' factors, and all the prior odds are also 1. The posterior densities of any quantity of interest ($\Delta$) existing in all the competing models can be combined together through BMA as \cite{kass_1995,hoeting_1999,fragoso_2018}:

\begin{dmath}
\label{eq 12}
p(\Delta|D)=\sum_{i=1}^{K} p(\Delta|D,M_i)p(M_i|D).
\end{dmath}

\subsection{Applied Error Correlation-based Model Fusion}

In the case of multiple uncertain sources of information (e.g., different models for the same problem), there is a need to integrate all the sources to produce more reliable results~\cite{dasey_2007}. In practice, there are several approaches for fusing information from multiple models. BMA is a model fusion technique that has some benefits in robust design. Other available techniques are fusion under known correlation~\cite{geisser_1965,morris_1977,winkler_1981,ghoreishi2018fusion}, and the covariance intersection method~\cite{julier_1997}. The key distinction of BMA over other model fusion approaches is the assumption of statistical independence among models, which may be incorrect in some cases and can lead to potentially serious misconceptions regarding confidence in quantity of interest estimates. 

A fundamental claim in this work is that \emph{any model can provide potentially useful information to a given task}. We thus seek to take into account all potential information any given model may provide and fuse unique information from the available models. Our fusion goal then is to identify dependencies, via estimated correlations, among the model discrepancies. With these estimated correlations, the models are fused following standard practice for the fusion of normally distributed data.

Under the case of known correlations between the discrepancies of models, the fused mean and variance at a design point, $\mathbf{x}$, are shown to be~\cite{winkler_1981}

\begin{dmath}
\label{eq 13}
E[\hat{f}(\mathbf{x})]=\frac{\mathbf{e}^\top \tilde{\Sigma}(\mathbf{x})^{-1}  {\mu}(\mathbf{x})}{\mathbf{e}^\top \tilde{\Sigma}(\mathbf{x})^{-1}  \mathbf{e}},
\end{dmath}

\begin{dmath}
\label{eq 14}
\textrm{Var}\left(\hat{f}(\mathbf{x})\right) = \frac{1}{\mathbf{e}^\top \tilde{\Sigma}(\mathbf{x})^{-1} \mathbf{e}},
\end{dmath} 
where $\mathbf{e} = [1,\ldots,1]^\top$, ${\mu}(\mathbf{x}) = [\mu_1(\mathbf{x}),\ldots,\mu_S(\mathbf{x})]^\top$ is the vector of mean values given $S$ models, and $\tilde{\Sigma}(\mathbf{x})^{-1}$ is the inverse of the covariance matrix between the models given as 

\begin{dmath}
\label{eq 15}
\tilde{\Sigma}=
\begin{bmatrix}
\sigma_1^2&\rho_{12}\,\sigma_1\sigma_2&\cdots&\rho_{1S}\,\sigma_1\sigma_S\\
\rho_{12}\,\sigma_1\sigma_2&\sigma_2^2&\cdots&\rho_{2S}\,\sigma_2\sigma_S\\
\vdots&\vdots& \ddots&\vdots\\
\rho_{1S}\,\sigma_1\sigma_S&\rho_{2S}\,\sigma_2\sigma_S&\cdots&\sigma_S^2\\
\end{bmatrix},
\end{dmath}
where $\rho_{ij}$ is the correlation coefficient between the deviations of models $i$ and $j$ at point $\mathbf{x}$, and $\sigma_i^2$ is the variance of model $i$ at point $\mathbf{x}$. 

To estimate the correlations between the model deviations in the case that they are unknown, we use the reification process defined in \cite{allaire_2012,thomison_2017}, which refers to the process of treating each model, in turn, as ground truth. This means that we assume the data generated by the reified model represents the true quantity of interest. These data are used to estimate the correlation between the errors of the different models. The process is repeated for each model. The detailed process of estimating the correlation between the errors of two models can be found in \cite{allaire_2012,thomison_2017}.

Following Thomison et al.~\cite{thomison_2017}, assuming model $i$ is chosen to reify, 
the correlation coefficients between the models $i$ and $j$, for $j = 1, \dots, i-1, i+1, \dots, S$, are given as:

\begin{dmath}
\rho_{ij}(\mathbf{x}) = \frac{\sigma_i^2(\mathbf{x})}{\sigma_i(\mathbf{x}) \sigma_j(\mathbf{x})} = \frac{\sigma_i(\mathbf{x})}{\sqrt{\left(\mu_i(\mathbf{x})-\mu_j(\mathbf{x})\right)^2 + \sigma_i^2(\mathbf{x})}},
\label{eq 16}
\end{dmath}
where $\mu_i(\mathbf{x})$ and $\mu_j(\mathbf{x})$ are the mean values of models $i$ and $j$ respectively at design point $\mathbf{x}$, and $\sigma_i^2(\mathbf{x})$ and $\sigma_j^2(\mathbf{x})$ are the variances at point $\mathbf{x}$. The first subscript under the correlation coefficient denotes which model has been reified. Since the only information we have regarding which model we believe to be more realistic is the variance of each, in addition to reifying model $i$ to estimate the correlation, we also reify information source $j$ and estimate $\rho_{ji}(\mathbf{x})$. We then estimate the correlation between the errors as the variance-weighted average of the two correlation coefficients as follows:

\begin{dmath}
\bar{\rho}_{ij}(\mathbf{x}) = \frac{\sigma_j^2(\mathbf{x})}{\sigma_i^2(\mathbf{x}) + \sigma_j^2(\mathbf{x})} \rho_{ij}(\mathbf{x}) + \frac{\sigma_i^2(\mathbf{x})}{\sigma_i^2(\mathbf{x})+\sigma_j^2(\mathbf{x})} \rho_{ji}(\mathbf{x}).
\label{eq 17}
\end{dmath}
These correlations can then be used to estimate the mean and variance of the fused estimate from Equations~\ref{eq 13} and \ref{eq 14}.  

\section{Results and Discussion}

In CALPHAD modeling, it is a hard task, even for CALPHAD experts, to know to what order the polynomial term in Equation~\ref{eq 2} should be expanded for each phase in the system to obtain the closest phase diagram to the available data after parameter optimization. For this reason, different models may be suggested for a system with different orders of polynomial expansions. In the context of Bayesian statistics, the best model among the selected candidates according to the available data can be identified through the calculation of the Bayes' factor based on Equation~\ref{eq 5}. 

Each of the models are probabilisticly calibrated against the available calculated data for the phase diagram using one million MCMC samples of the parameter vectors. For each model, the deterministically optimized values of the parameters obtained from the PARROT module in Themo-Calc software (through least squares) and $\pm$ 3 times these values are considered as the initial values ($\theta_0$) and the ranges of the parameters, respectively. This prior information enables much faster convergence of the model parameters during MCMC sampling. However, it is assumed that the deterministic optimization in Thermo-Calc is global, which is not necessarily true as we discuss later. It should be noted that the compositions of stable phases estimated from the convex hull of the phase Gibbs energies at any given temperature are compared with the corresponding available data through the likelihood during the sampling process. Here, the results obtained for model 2 are discussed in detail to show how MCMC is used in this work for probabilistic calibration of the model parameters. The same approach is applied for the calibration of the other three model parameter sets.

\begin{figure*}[htp]
\centering
\begin{minipage}{0.45\textwidth}
  \centering
  \includegraphics[width=1\linewidth]{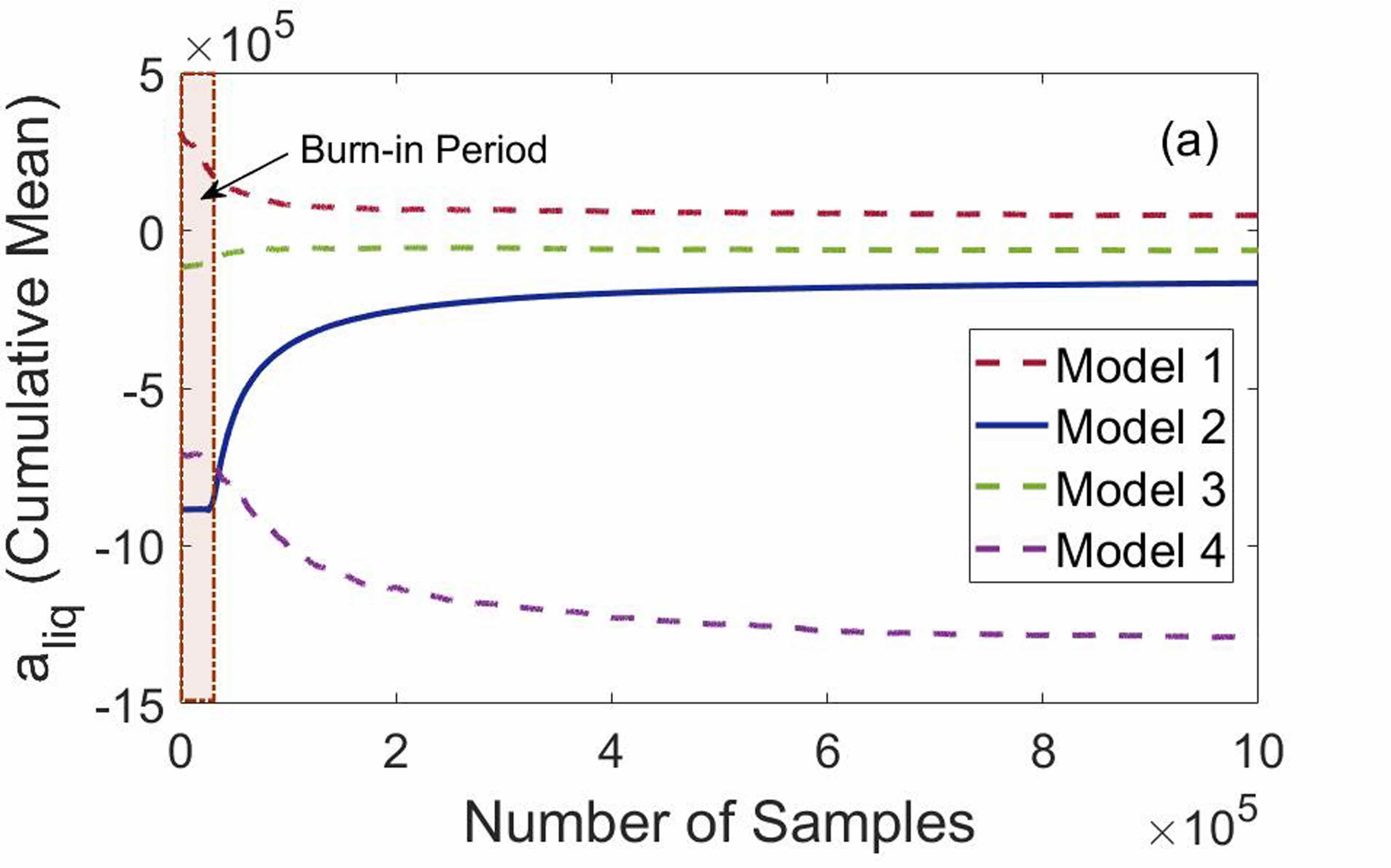}
  \label{Fig 1a}
\end{minipage}
\begin{minipage}{0.45\textwidth}
  \centering
  \includegraphics[width=1\linewidth]{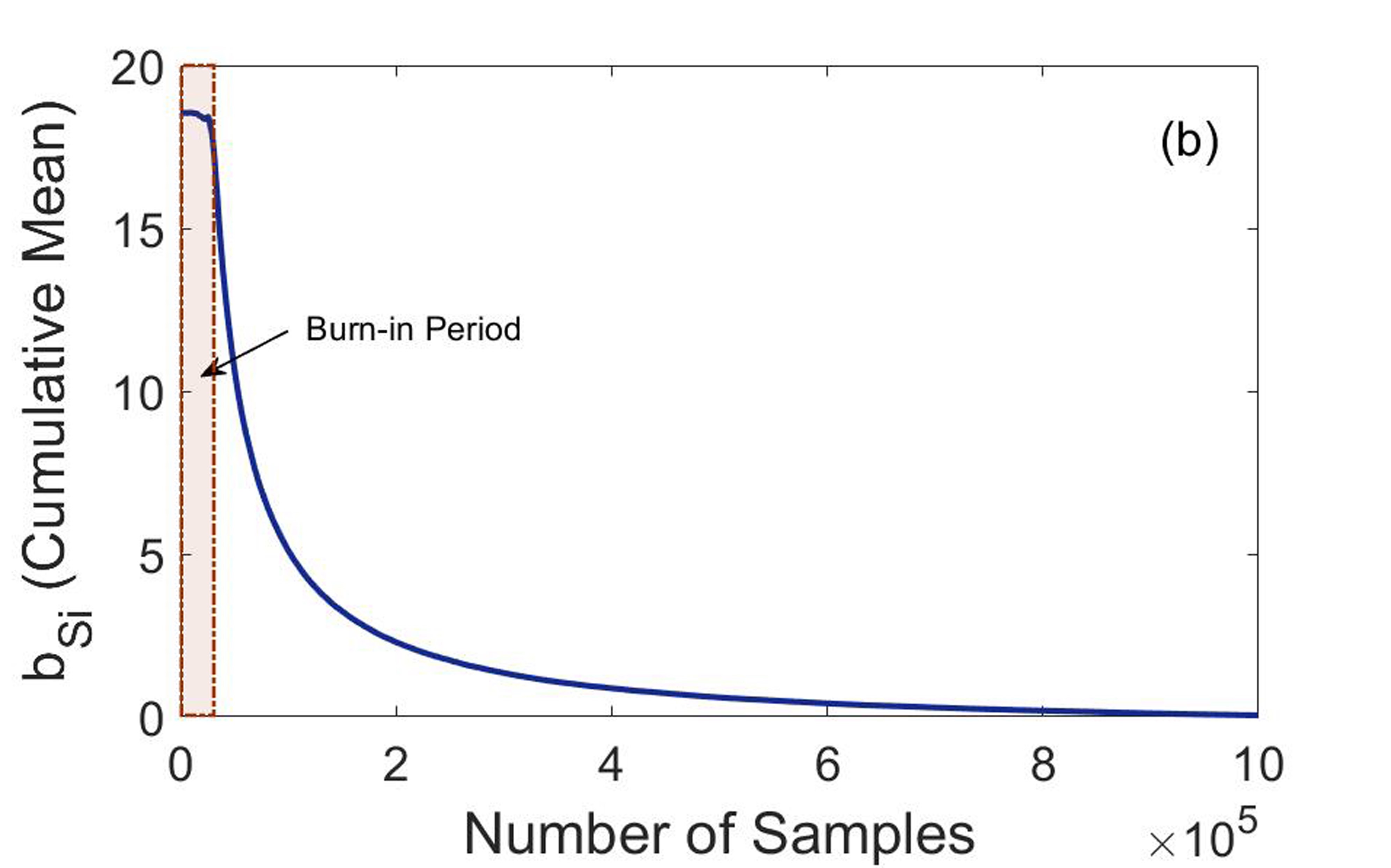}
\label{Fig 1b}
\end{minipage}
\caption{Cumulative mean plots of the parameters a) $a_{liq}$ and b) $b_{Si}$ in model 2. The red shaded regions show the number of parameters belonging to the burn-in period.} 
\label{Fig 1}
\end{figure*}

During MCMC sampling, there are some homogeneous fluctuations around plausible optimum values of parameters after convergence. However, this means the cumulative means of parameter samples should converge to constant values. Therefore, plotting cumulative mean distributions of parameters can help identify the convergence regions. As examples, two of these plots are shown in Figure~\ref{Fig 1}. According to this figure, smooth changes towards constant values are observed for both parameters after 30,000 parameter generations, which correspond to the parameter convergence regions. Therefore, the first 30,000 generated samples are considered as the burn-in period (red shaded regions) and discarded from the total number of parameter samples. Since all the parameters generally converge at the same time, the burn-in period is assumed to be the same for the other model parameters (the other 18 parameters in model 2). To show that the parameters in other models also converge to their optimum values, the cumulative means of parameter $a_{liq}$ are plotted for the other models as dotted lines in Figure~\ref{Fig 1}a.


After the removal of the burn-in period, the mean values of the remaining parameter samples (970,000 samples) and the diagonal terms of their variance-covariance matrix are considered as the optimal plausible mean values and the overall reflected uncertainties (variances) for the model parameters, respectively. The initial values of the model parameters obtained through a deterministic optimization and their probabilistic values after MCMC calibration are listed in Table~\ref{Table 2}.

\begin{table*}[htp]

\centering
\caption{Initial values for the parameters in model 2 and their plausible optimum values and overall uncertainties after MCMC calibration.}

\setlength{\arrayrulewidth}{1mm}
\setlength{\tabcolsep}{12pt}
\renewcommand{\arraystretch}{1.2}
\scalebox{0.55}
{\begin{tabular}{c c c c}

\hline\hline

\textbf{Phase} & \textbf{Parameter Specification} & \textbf{Deterministically Calibrated Initial Values} & \textbf{Probabilistically Calibrated Values after MCMC} \\

\hline

$HCP-Si$ & ${^{0}{L^{HCP}_{Hf,Si}}: {^{0}{a^{HCP}_{Hf,Si}}} - {^{0}{b^{HCP}_{Hf,Si}}}}T$ & ${^{0}{L^{HCP}_{Hf,Si}}}: - 455420 + 221.79T$ & ${^{0}{L^{HCP}_{Hf,Si}}}: - 42156\pm22343 + 21.55\pm11.04T$ \\

$BCC-Si$ & ${^{0}{L^{BCC}_{Hf,Si}}: {^{0}{a^{BCC}_{Hf,Si}}} - {^{0}{b^{BCC}_{Hf,Si}}}}T$ & ${^{0}{L^{BCC}_{Hf,Si}}}: - 45254 + 18.56T$ & ${^{0}{L^{BCC}_{Hf,Si}}}: 2307\pm1435 - 0.50\pm0.58T$ \\

$Hf_2Si$ & ${^{0}{L^{Hf_2Si}_{Hf,Si}}: {^{0}{a^{Hf_2Si}_{Hf,Si}}} - {^{0}{b^{Hf_2Si}_{Hf,Si}}}}T$ & ${^{0}{L^{Hf_2Si}_{Hf,Si}}}: -454090 + 90.46T$ & ${^{0}{L^{Hf_2Si}_{Hf,Si}}}: - 45513\pm12625 - 0.11\pm2.91T$ \\

$Hf_3Si_2$ & ${^{0}{L^{Hf_3Si_2}_{Hf,Si}}: {^{0}{a^{Hf_3Si_2}_{Hf,Si}}} - {^{0}{b^{Hf_3Si_2}_{Hf,Si}}}}T$ & ${^{0}{L^{Hf_3Si_2}_{Hf,Si}}}: - 481270 + 84.19T$ & ${^{0}{L^{Hf_3Si_2}_{Hf,Si}}}: - 45445\pm13335 - 3.54\pm2.79T$ \\

$Hf_5Si_3$ & ${^{0}{L^{Hf_5Si_3}_{Hf,Si}}: {^{0}{a^{Hf_5Si_3}_{Hf,Si}}} - {^{0}{b^{Hf_5Si_3}_{Hf,Si}}}}T$ & ${^{0}{L^{Hf_5Si_3}_{Hf,Si}}}: - 454950 + 77.55T$ & ${^{0}{L^{Hf_5Si_3}_{Hf,Si}}}: - 41728\pm12666 - 4.26\pm2.62T$ \\

$Hf_5Si_4$ & ${^{0}{L^{Hf_5Si_4}_{Hf,Si}}: {^{0}{a^{Hf_5Si_4}_{Hf,Si}}} - {^{0}{b^{Hf_5Si_4}_{Hf,Si}}}}T$ & ${^{0}{L^{Hf_5Si_4}_{Hf,Si}}}: - 522680 + 100.88T$ & ${^{0}{L^{Hf_5Si_4}_{Hf,Si}}}: - 47201\pm14009 - 3.06\pm3.08T$ \\

$HfSi$ & ${^{0}{L^{HfSi}_{Hf,Si}}: {^{0}{a^{HfSi}_{Hf,Si}}} - {^{0}{b^{HfSi}_{Hf,Si}}}}T$ & ${^{0}{L^{HfSi}_{Hf,Si}}}: - 503250 + 98.26T$ & ${^{0}{L^{HfSi}_{Hf,Si}}}: - 49654\pm14110 - 1.63\pm3.28T$ \\

$HfSi_2$ & ${^{0}{L^{HfSi_2}_{Hf,Si}}: {^{0}{a^{HfSi_2}_{Hf,Si}}} - {^{0}{b^{HfSi_2}_{Hf,Si}}}}T$ & ${^{0}{L^{HfSi_2}_{Hf,Si}}}: - 468950 + 132.66T$ & ${^{0}{L^{HfSi_2}_{Hf,Si}}}: - 48537\pm13301 + 5.42\pm4.23T$ \\

$liquid$ & ${^{0}{L^{liq}_{Hf,Si}}: {^{0}{a^{liq}_{Hf,Si}}} - {^{0}{b^{liq}_{Hf,Si}}}}T$ & ${^{0}{L^{liq}_{Hf,Si}}}: - 886370 - 9.47T$ & ${^{0}{L^{liq}_{Hf,Si}}}: - 145597\pm30285 + 15.56\pm3.04T$ \\

$Si$ & ${^{0}{L^{Si}_{Hf,Si}}: {^{0}{a^{Si}_{Hf,Si}}} - {^{0}{b^{Si}_{Hf,Si}}}}T$ & ${^{0}{L^{Si}_{Hf,Si}}}: - 114810 + 68.22T$ & ${^{0}{L^{Si}_{Hf,Si}}}: 6388\pm3600 - 3.77\pm2.14T$ \\

\hline\hline

\end{tabular}}

\label{Table 2}
\end{table*}

\begin{figure*}[htp]
\centering
\begin{minipage}{0.45\textwidth}
  \centering
  \includegraphics[width=1\linewidth]{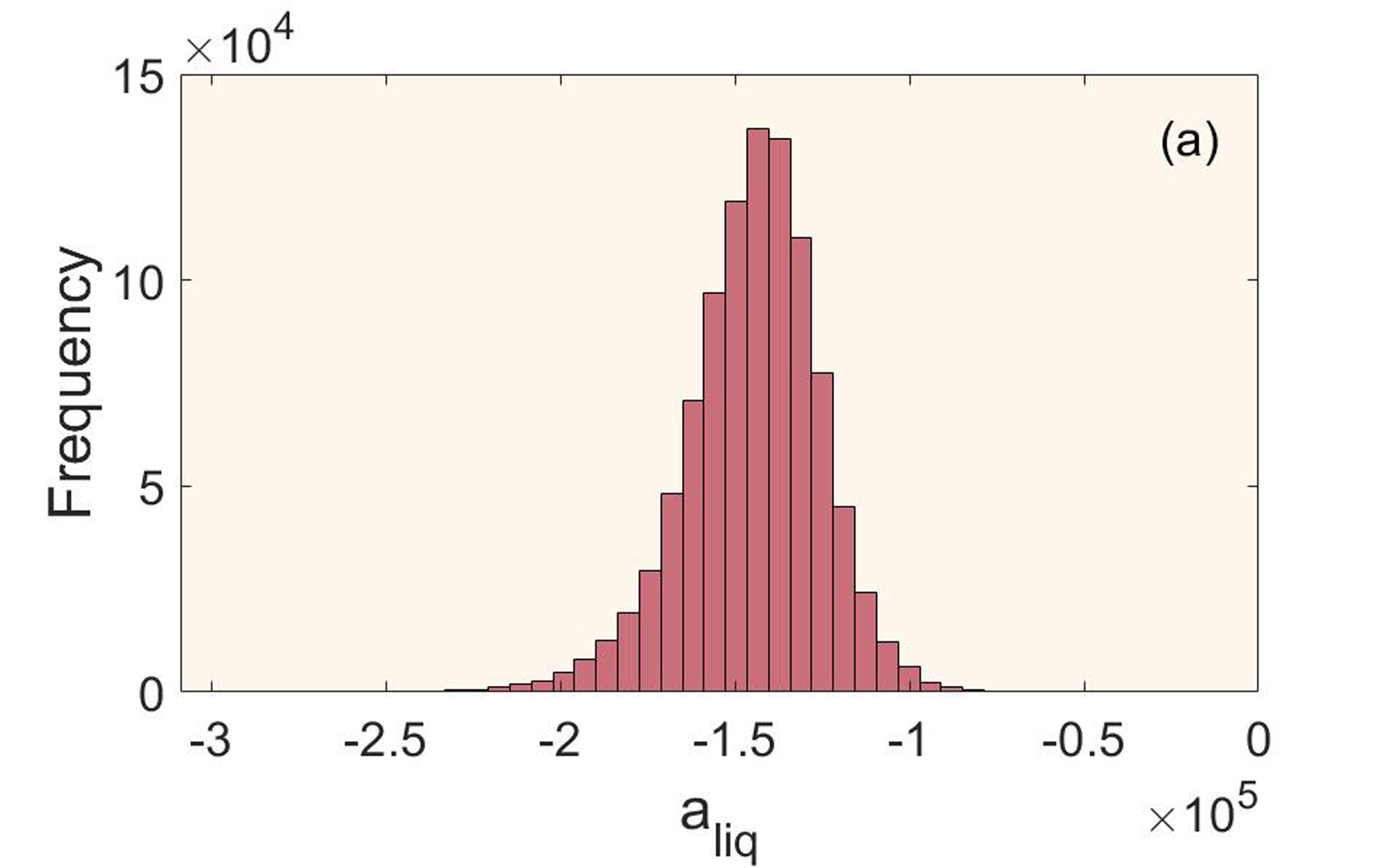}
  \label{Fig 2a}
\end{minipage}
\begin{minipage}{0.45\textwidth}
  \centering
  \includegraphics[width=1\linewidth]{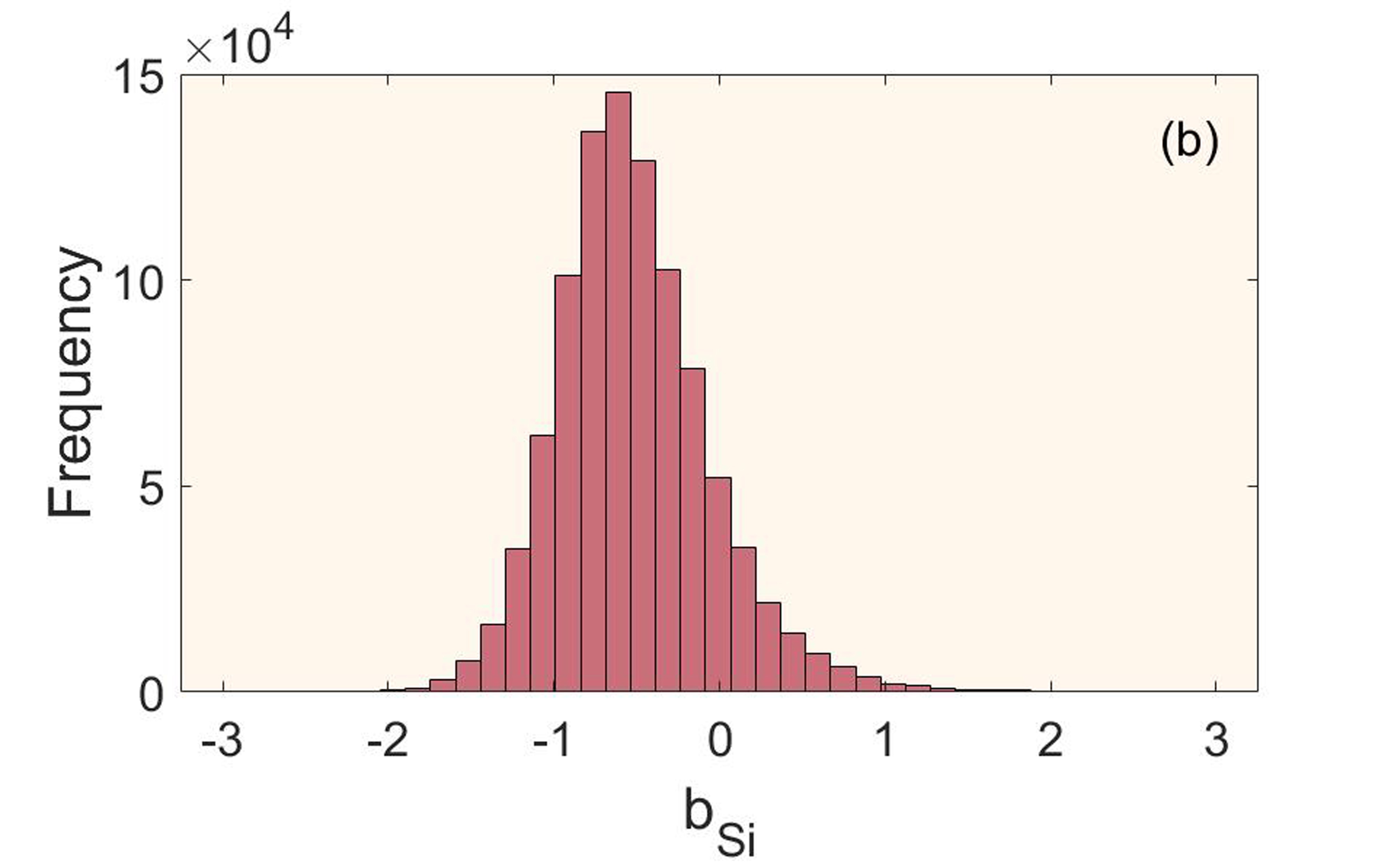}
\label{Fig 2b}
\end{minipage}
\caption{Marginal posterior frequency distributions for the parameters a) $a_{liq}$ and b) $b_{Si}$ in model 2.\label{Fig 2}} 
\end{figure*}

\begin{figure*}[htp]
\centering
\begin{minipage}{0.45\textwidth}
  \centering
  \includegraphics[width=1\linewidth]{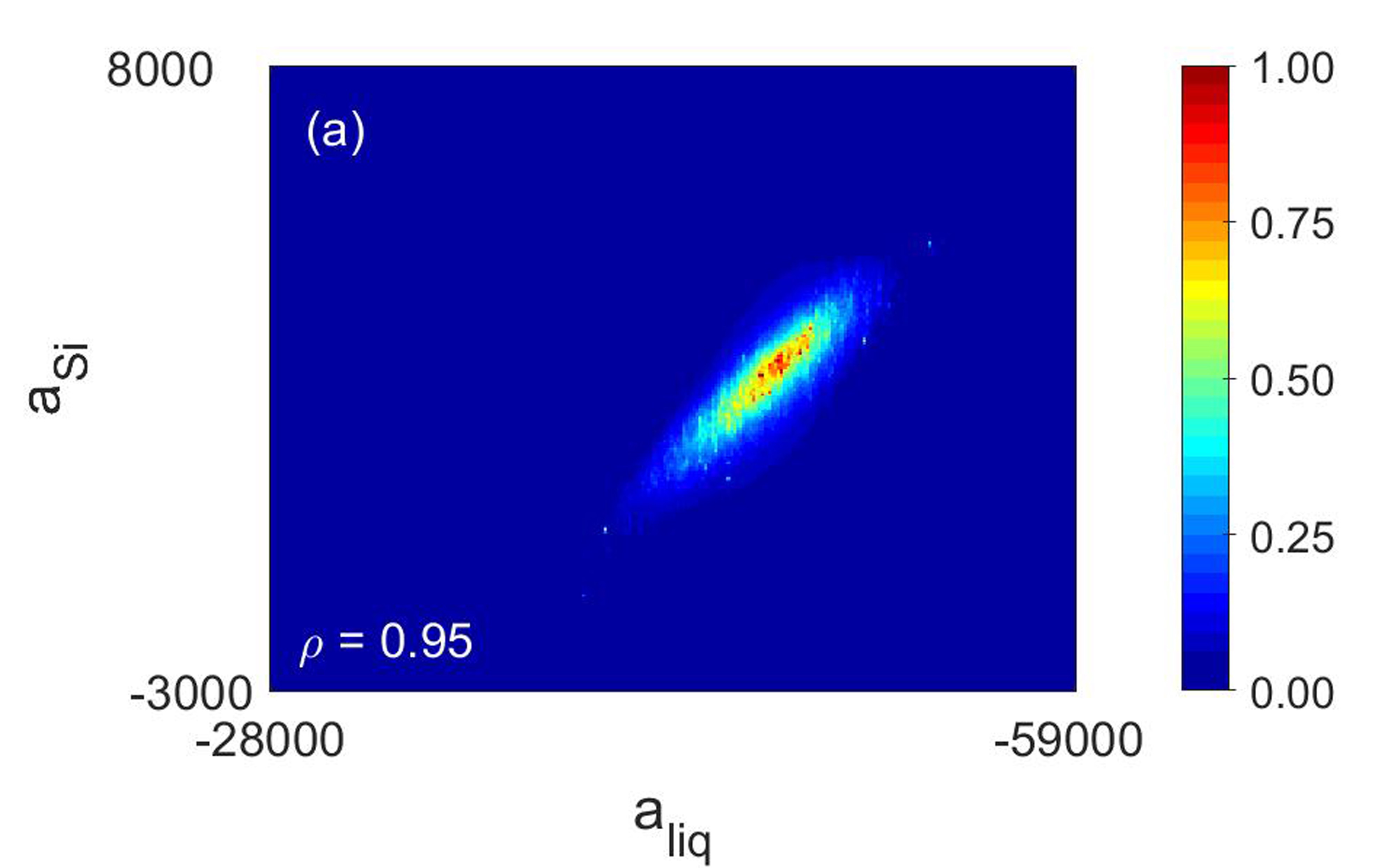}
  \label{Fig 3a}
\end{minipage}
\begin{minipage}{0.45\textwidth}
  \centering
  \includegraphics[width=1\linewidth]{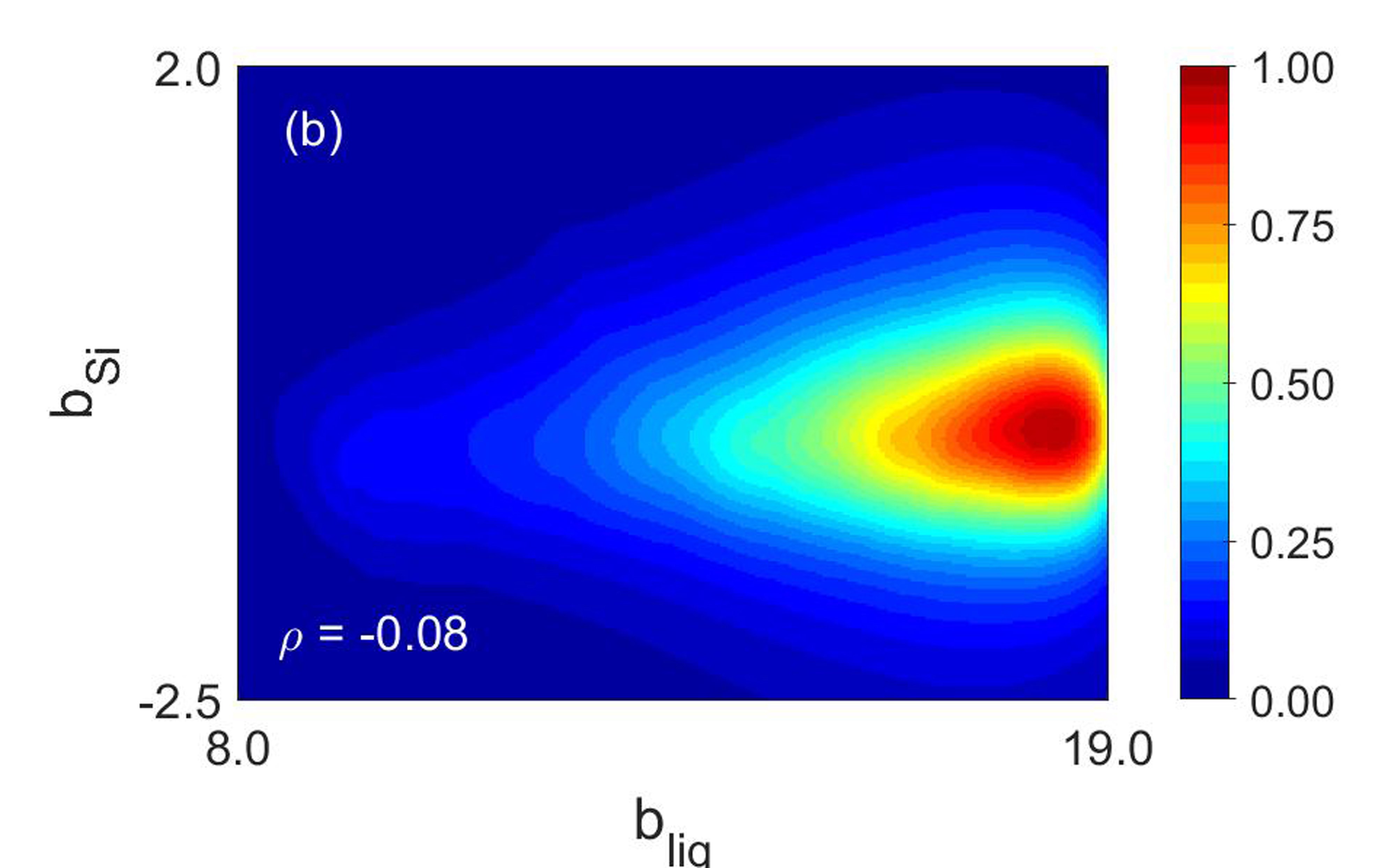}
\label{Fig 3b}
\end{minipage}
\caption{Joint frequency distributions between the pair parameters a) ($a_{Si}$,$a_{liq}$) and b) ($b_{Si}$,$b_{liq}$) in model 2 besides their linear correlation coefficients and normalized colour bars.\label{Fig 3}} 
\end{figure*}

The marginal and joint posterior frequency distributions of the parameters can also be plotted to evaluate the convergence and dispersion of samples in the parameter space. Two examples of marginal posterior frequency histograms for parameters are plotted in Figure~\ref{Fig 2}. As can be observed in this figure, the parameter posteriors are almost in the form of Gaussian distributions. Since all the parameters show a similar form for their marginal posterior distributions, a distribution close to a multivariate Gaussian distribution is expected as the joint distribution for all the model parameters (20 parameters in the case of model 2). Figure \ref{Fig 3} shows the joint frequency distributions for two examples of a pair of parameters in model 2. Figure \ref{Fig 3} can also offer a qualitative representation of the correlation between the applied two parameters in each case. For example, the linearity and direction of the red regions show the strength and negativity/positivity of the correlation between parameters, quantified through the Pearson coefficient\cite{battle_2004}:

\begin{equation}
\label{eq 18}
\rho_{x,y}=\frac{cov(x,y)}{\sigma_x \sigma_y}
\end{equation}


\begin{figure*}[htp]
\centering
\begin{minipage}{0.45\textwidth}
  \centering
  \includegraphics[width=1\linewidth]{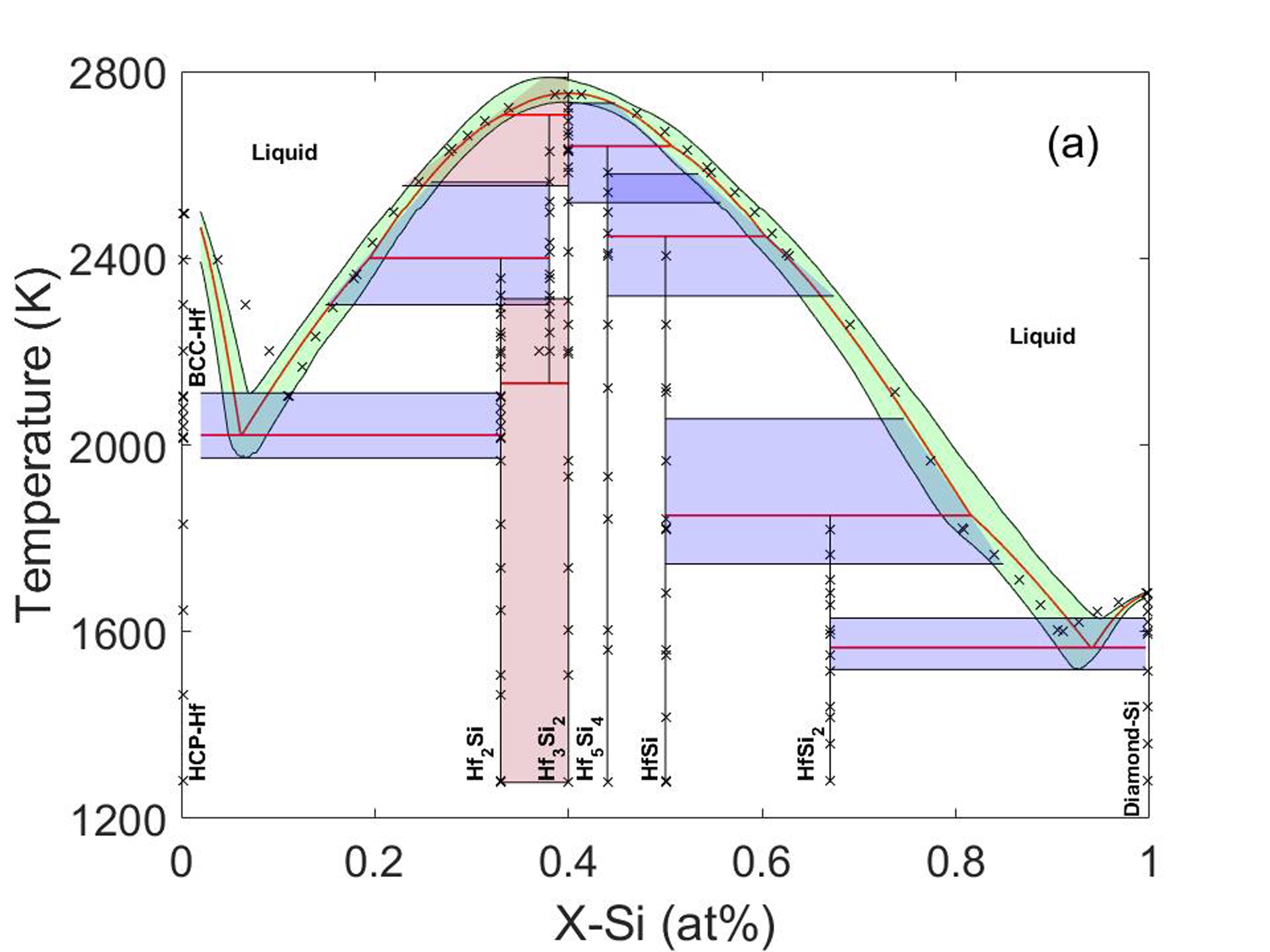}
  \label{Fig 4a}
\end{minipage}
\begin{minipage}{0.45\textwidth}
  \centering
  \includegraphics[width=1\linewidth]{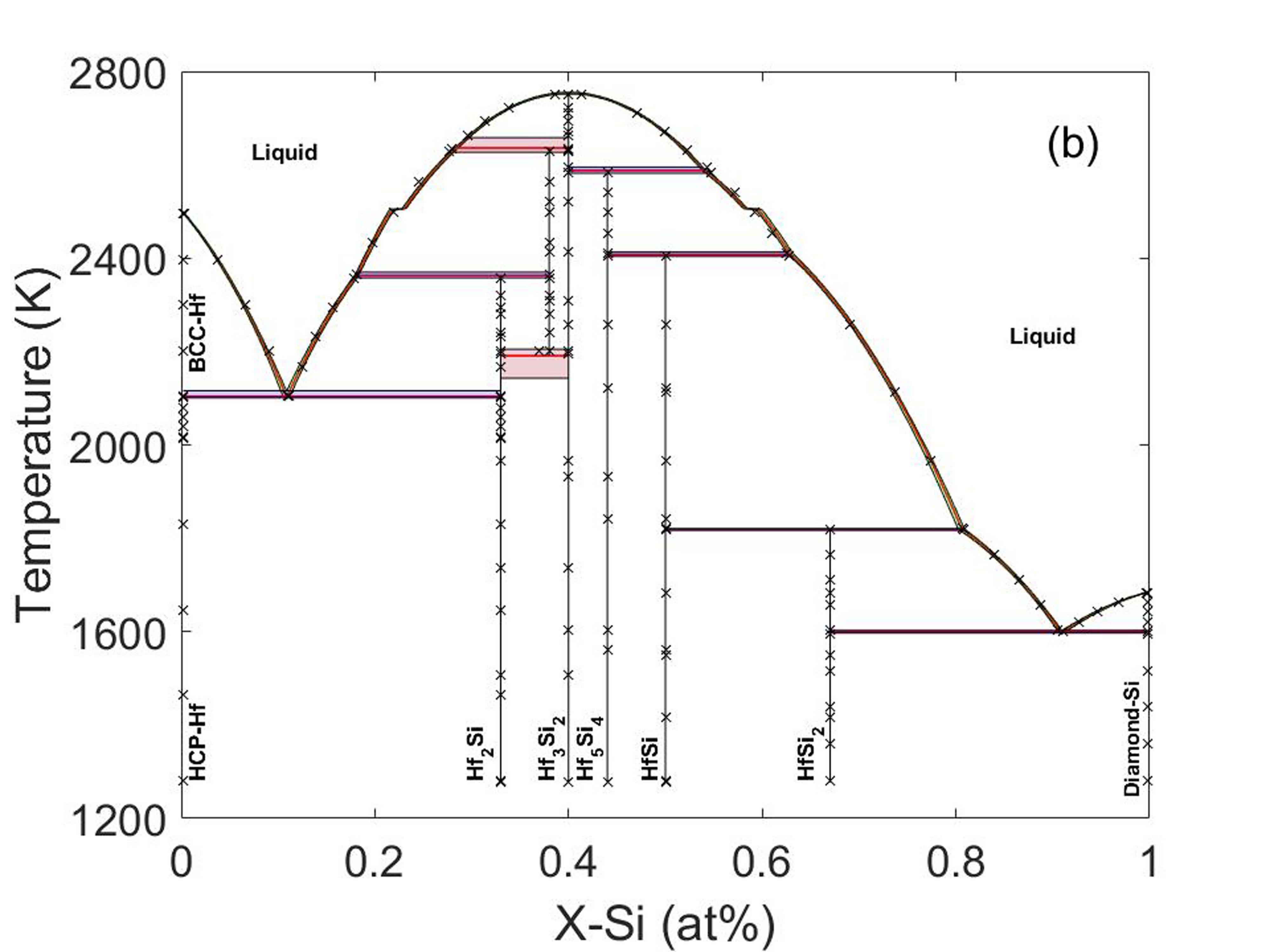}
\label{Fig 4b}
\end{minipage}
\begin{minipage}{0.45\textwidth}
  \centering
  \includegraphics[width=1\linewidth]{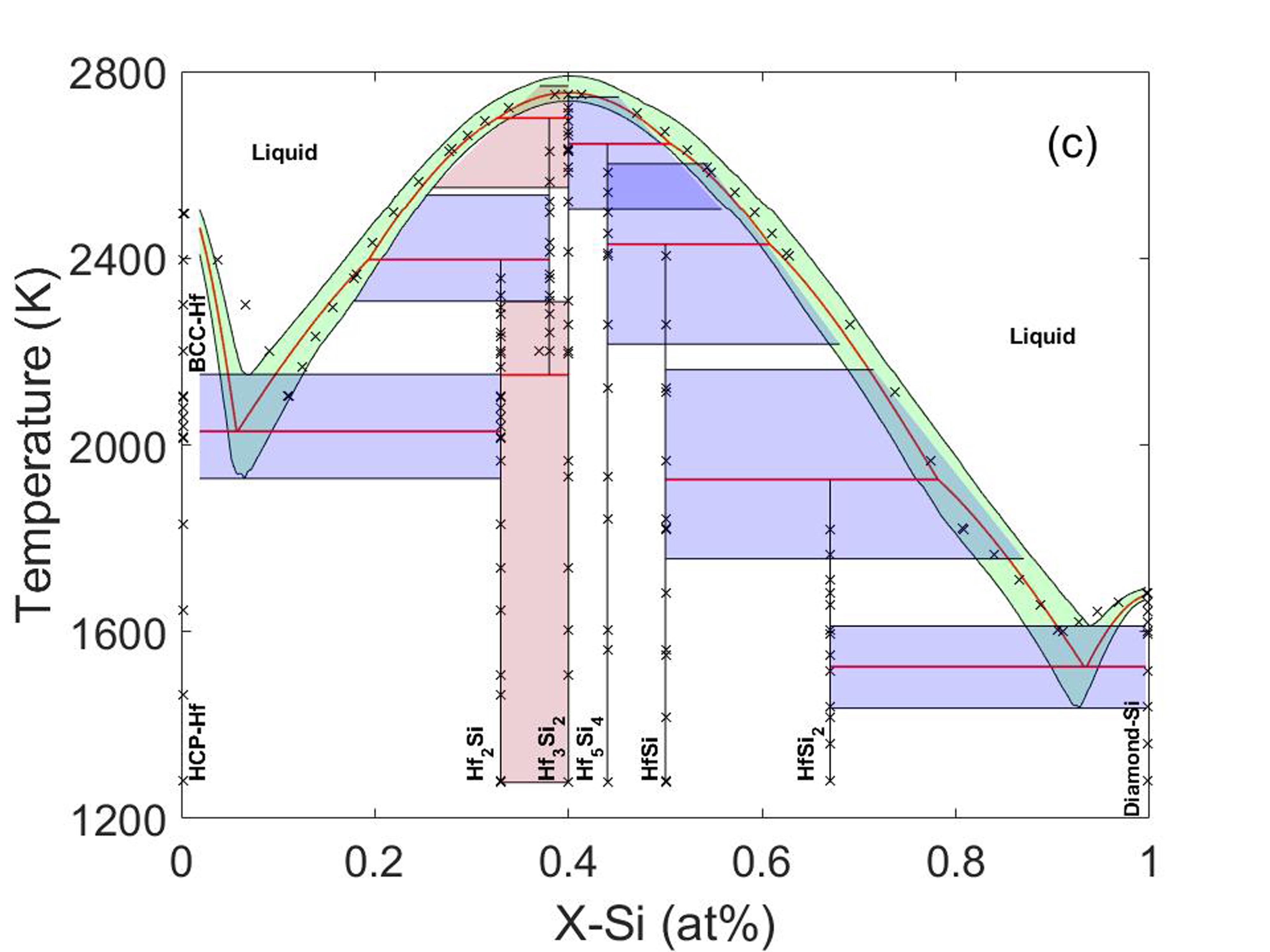}
\label{Fig 4c}
\end{minipage}
\begin{minipage}{0.45\textwidth}
  \centering
  \includegraphics[width=1\linewidth]{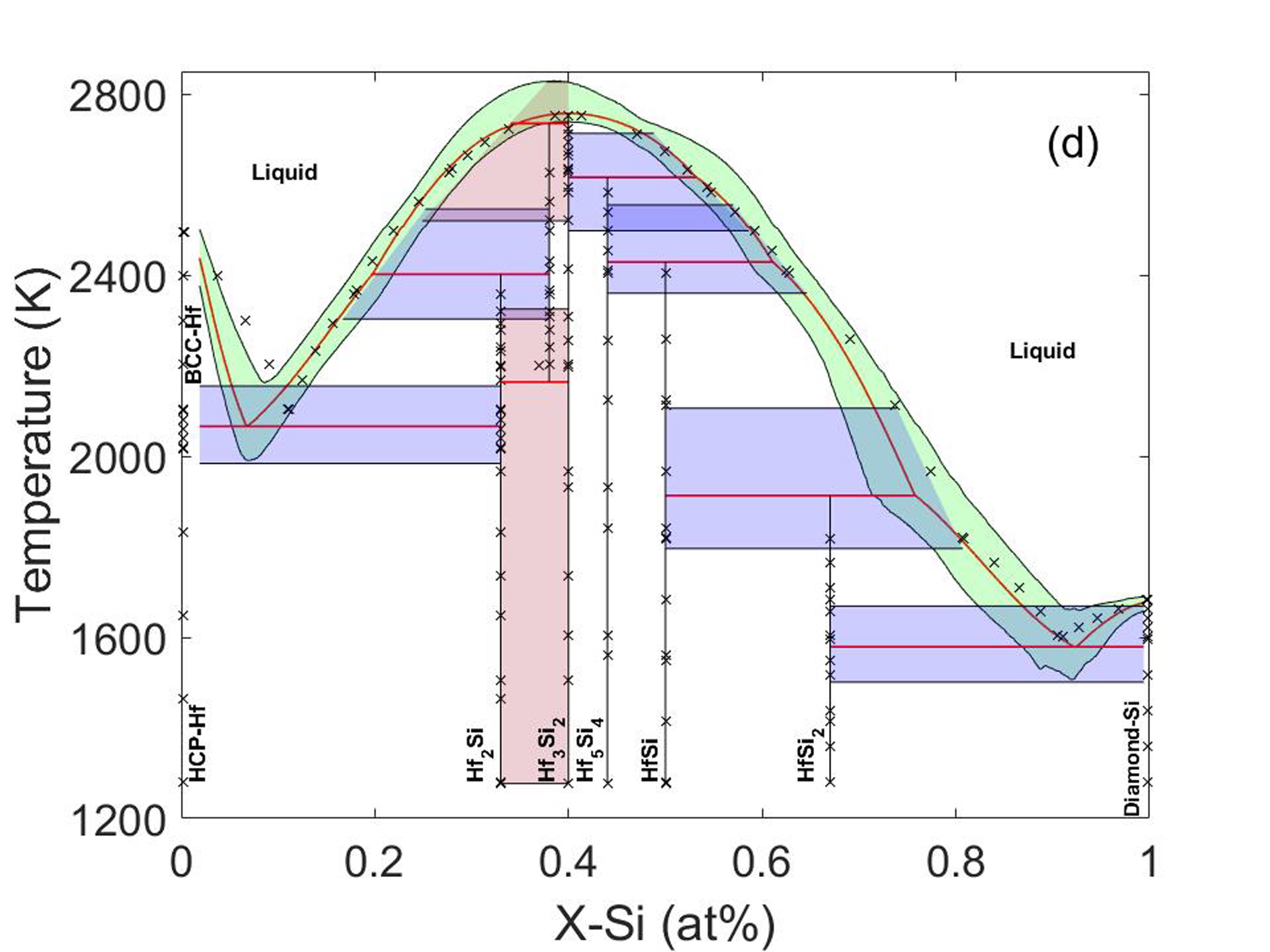}
\label{Fig 4d}
\end{minipage}
\caption{Optimum Hf-Si phase diagrams and their 95\% BCIs obtained from models 1-4 (a-d) after uncertainty propagation of the MCMC calibrated parameters in each case.\label{Fig 4}}
\end{figure*}

The same MCMC approach is applied to probabilistically calibrate the other three model parameter sets. Then, the overall uncertainties are propagated from parameters to phase diagram for each model through the model forward analysis. This process is performed using the forward calculation of the phase diagram by the last 5,000 MCMC parameter samples as an ensemble of the whole convergence region. To find 95\% Bayesian credible intervals (BCI), 2.5\% of the resulting samples associated with liquidus and transformation lines are discarded from above and also below the total obtained uncertainty band at any specified composition. These results are shown for each model in Figure~\ref{Fig 4}. In this figure, red lines and shaded areas are the results obtained from the optimum (mean) values of the parameters and 95\% BCI for each model, respectively. A cross-sectional probability distribution can be achieved at any specific Si content along the green uncertainty interval, while there is a fixed (composition independent) cross-sectional probability distribution throughout any one of the blue or red uncertainty intervals. The red shaded regions in all models show unstable results with large uncertainties that result from the difficulty to predict the intermediate high temperature Hf$_5$Si$_3$ phase through CALPHAD modeling. Moreover, it is clear in Figure~\ref{Fig 4} that there is a very good agreement between the results obtained from model 2 and the data with a very small uncertainty band. It should be noted that indentations around $X_{Si}=0.2$ and $0.6$ at 2500K are caused by numerical errors in the optimization process.


In the context of BMS/BMA, the weight of each one of the applied models can be obtained by calculating their corresponding posteriors given data using Equations~\ref{eq 10} and \ref{eq 11}. Based on these calculations, the weights associated with models 1 to 4 are 0.1474, 0.5572, 0.1451, and 0.1504, respectively. Model 2 thus has three times the weight of the other models, which otherwise have similar Bayesian importance, which is consistent with the phase diagram results in Figure~\ref{Fig 4}.

Traditionally, in the CALPHAD community,  it has been generally expected that a better fit to experimental data can be achieved by simply increasing the number of parameters in the Redlich-Kister polynomial expansions of the free energies. However, our results show that this is not always the case. The results hold, for example, when comparing model 1 to model 2, as increasing the complexity of the free energy functions clearly results in a much better fit to the data and a much narrower uncertainty bound. As the complexity of the models are increased, however, it is apparent that the uncertainty bounds become much worse. In this case, we have found that the most parsimonious model (i.e., simplest) that still has sufficient freedom to explain all the available data is clearly superior.

\begin{figure*}[htp]
\centering
  \includegraphics[width=0.7\linewidth]{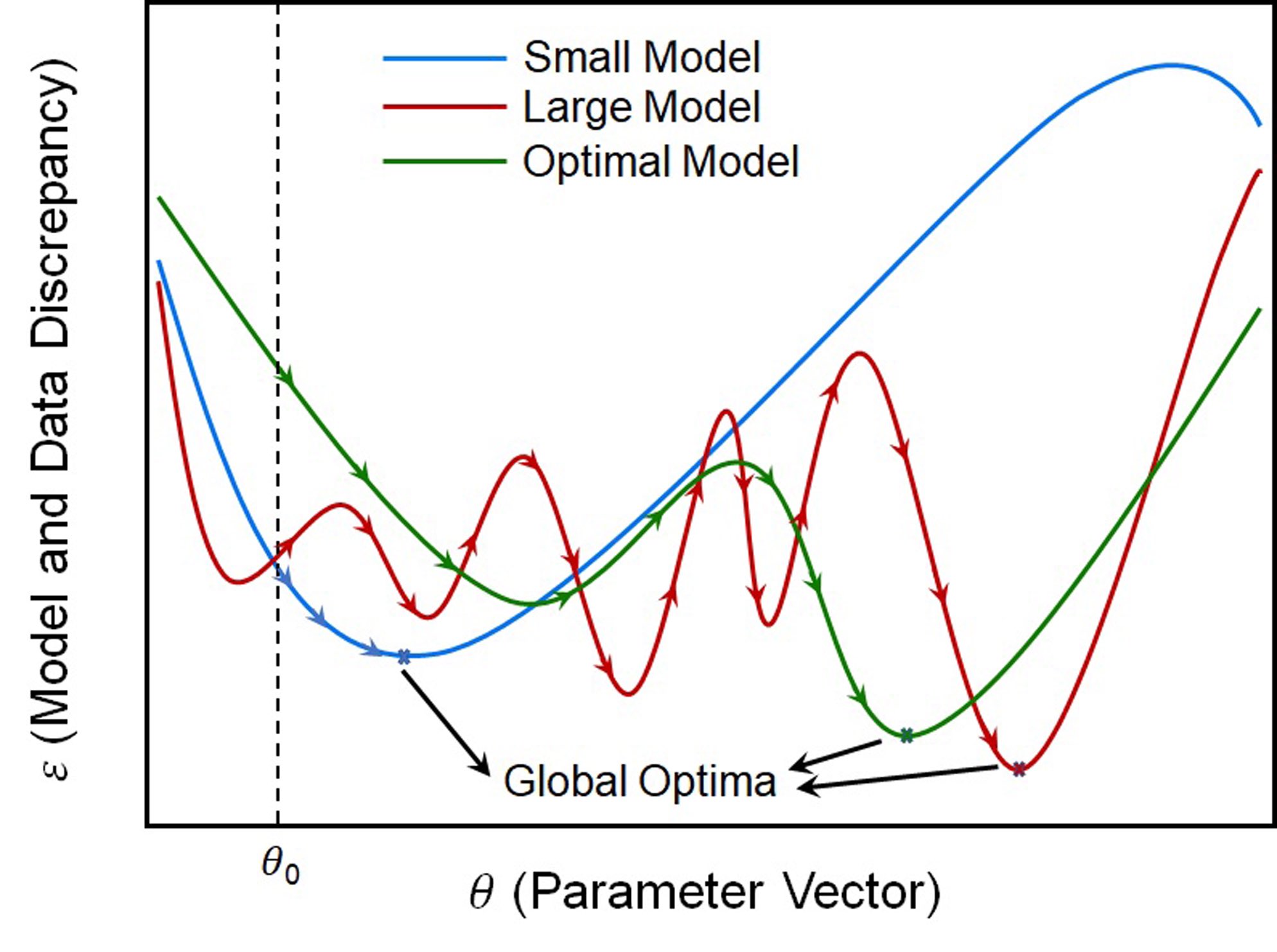}
\caption{2D schematic plots of the discrepancy between model and data in the parameter space for different model sizes (different number of parameters).\label{Fig 5}}
\end{figure*}

These results, although counter-intuitive to a degree, may be explained by imagining the fitness landscape of a particular model as a multi-dimensional space, with many local minima corresponding to combinations of parameters that result in a fit with lower error relative to the neighborhood in model parameter space. Schematically, one can visualize this as shown in Figure \ref{Fig 5}. The simplest model (blue) is depicted as having a relatively simple landscape with a sub-optimal fitness: the model is too simple to explain all the available data and the large uncertainty bounds are associated with the fitness error as a result of (small) variations in the values of individual parameters. Very complex models (red) have a high number of local minima as there are many combinations of parameters that result in somewhat-optimal fits. The fitting process, however, can get stuck in these local minima and since the MCMC results are inevitably biased by the parameter ranges defined based on the deterministically-attained values---to sample the model space at a reasonable cost---there is a significant probability that the global minimum is outside the parameter ranges, which results in an MCMC chain that converges to a sub-optimal region during MCMC sampling. The large uncertainty bounds in this case could be ascribed to the large sensitivity of such models to small deviations in the values of the parameters. The best model (green) has the right amount of complexity to fit all the available data without having too much model parameter uncertainty. There is thus a valid argument towards \emph{model parsimony}: to make the model as simple as possible, but not more.


\begin{figure*}[htp]
\centering
  \includegraphics[width=1\linewidth]{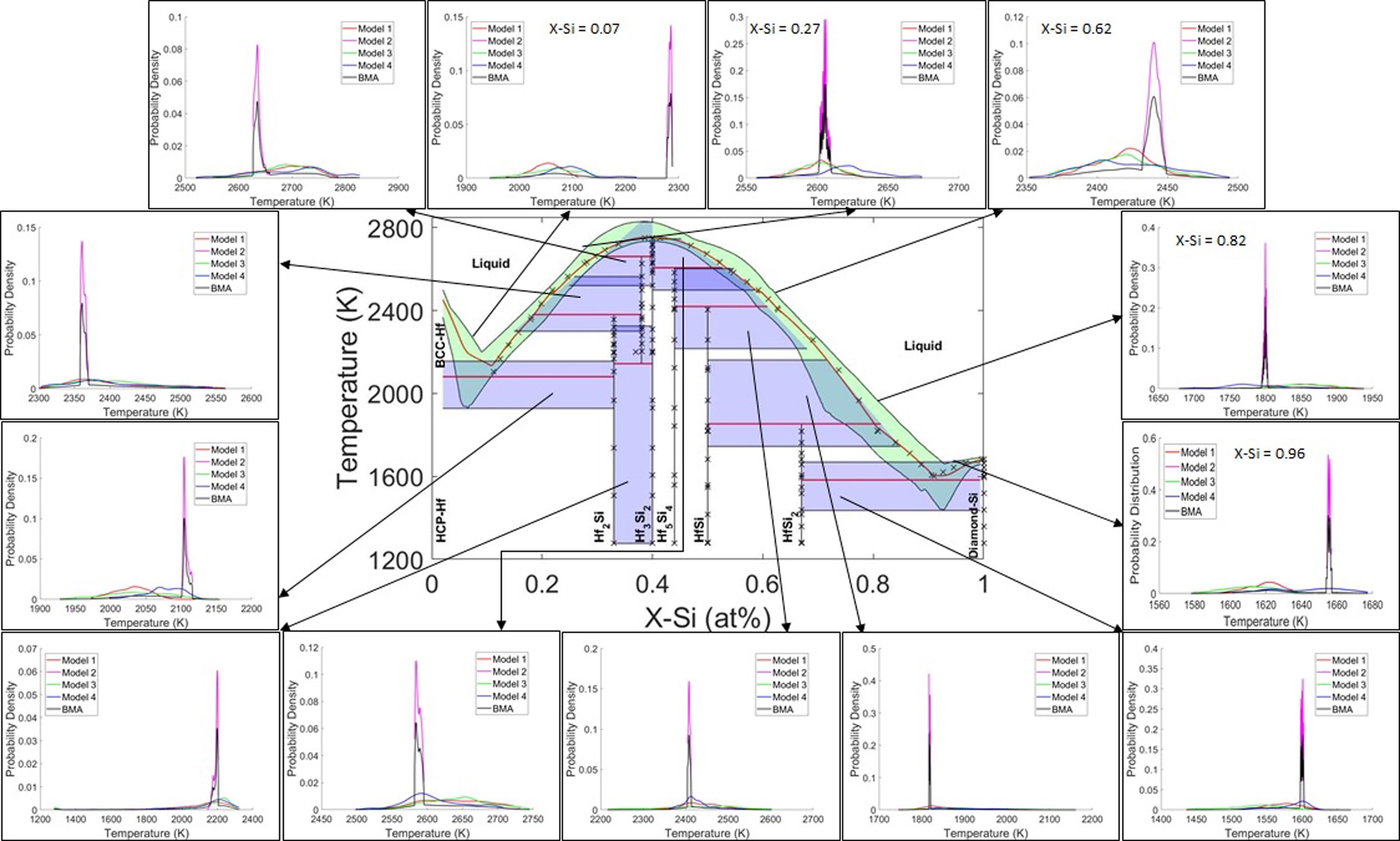}
\caption{Mean values and 95\% BCIs at different compositions/regions in Hf-Si phase diagram obtained after BMA.\label{Fig 6}}
\end{figure*}

It must be stated, however, that while there is a clear superiority in model 2 in this case, the value of the inferior models should not be discounted outright as they may provide valuable information that can contribute to a more robust quantification of uncertainty. Based on Equation~\ref{eq 12}, weighted averages are taken over the cross-sectional probability distributions of all four phase diagrams in Figure~\ref{Fig 4} at different specified compositions/regions to calculate the average model and its uncertainty bands throughout the whole phase diagram, as shown in Figure~\ref{Fig 6}. For each individual model and weighted average model, the cross-sectional probability distributions obtained through kernel density estimation (KDE) are also demonstrated for all the transformation line uncertainty bands (blue shaded regions with composition independent distributions) and at some random compositions for the liquidus uncertainty band (green shaded region) in  Figure~\ref{Fig 6}. As can be observed, 95\% BCIs of the average model (black lines in probability distribution plots) is broader than each one of the applied models and covers all the uncertainty ranges offered by the models 1 to 4. This is why BMA can provide more confidence in the results in the context of robust design.

\begin{figure*}[htp]
\centering
  \includegraphics[width=0.7\linewidth]{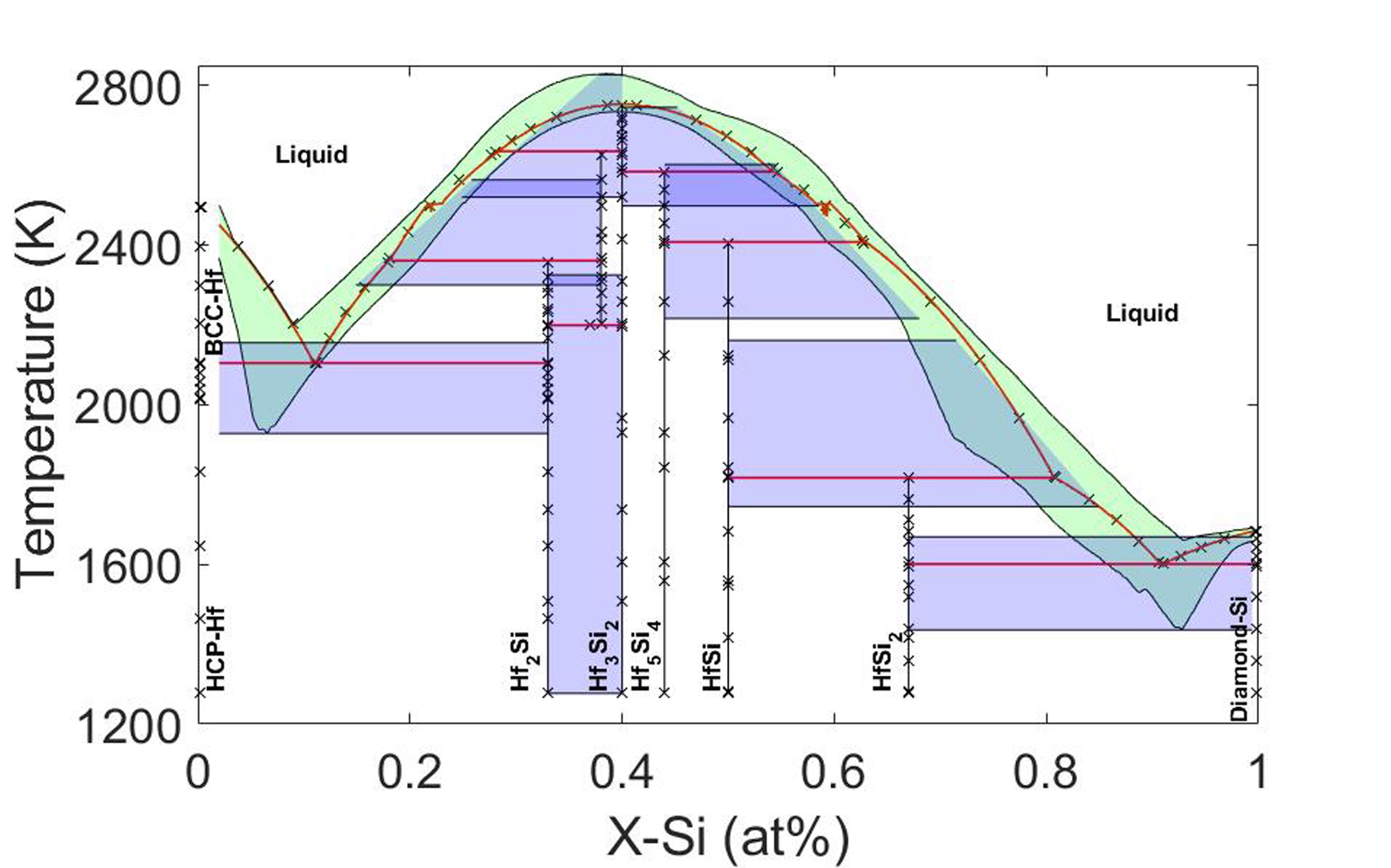}
\caption{Posterior modes and 95\% BCIs at different compositions/regions in Hf-Si phase diagram obtained after BMA.\label{Fig 7}}
\end{figure*}

As observed in Figure~\ref{Fig 6}, the phase diagram resulting from the mean value of probability distributions at different compositions and regions (red lines) leads to two unrealistic results (indentations) around the eutectic points at approximately $X_{Si}=0.1$ and $0.9$. According to Figure~\ref{Fig 6}, this issue can be attributed to the difference of models in the prediction of eutectic point compositions, which results in averages of some hypo-eutectic points from model 2 with some hyper-eutectic points from models 1, 3 and 4 at small composition ranges. In order to solve this issue, the posterior modes of the probability distributions can be introduced as the optimum phase diagram rather than their mean values. The result is shown by red lines in Figure~\ref{Fig 7}. It is worth noting that the posterior mode of the probability distributions in the average model exactly corresponds to the posterior modes of the probability distributions in model 2, as indicated by black and purple lines in distribution plots of Figure~\ref{Fig 7}, respectively. Therefore, the best model results can be considered as the optimum results for the average model, but with broader uncertainties, contributed by the inferior models.


Generally, the dependency of the model precision to the number of parameters defined for the existing phases in the system has made CALPHAD modeling a hard task that requires a lot of experience. Therefore, the lack of a systematic approach towards thermodynamic assessments means that model search is essentially carried out in a trial-and-error mode or by relying on (considerable) expert opinion. While the BMA approach already put forward in this work provides a robust estimate of the (weighed) output of a set of models, one drawback is that BMA assumes that models are statistically independent from each other. In this specific case, since the three suboptimal models have almost as much weight at the optimal model, the uncertainties carried out by the former over-estimate (in a very conservative manner) the fused uncertainty. All models (at least in this case) have some degree of correlation as (i) they are describing the same underlying ground truth; (ii) have common model structures. We thus propose to exploit the statistical correlations between models as a strategy to arrive at an improved fused prediction of phase stability. 

\begin{figure*}[htp]
\centering
\begin{minipage}{0.45\textwidth}
  \centering
  \includegraphics[width=1\linewidth]{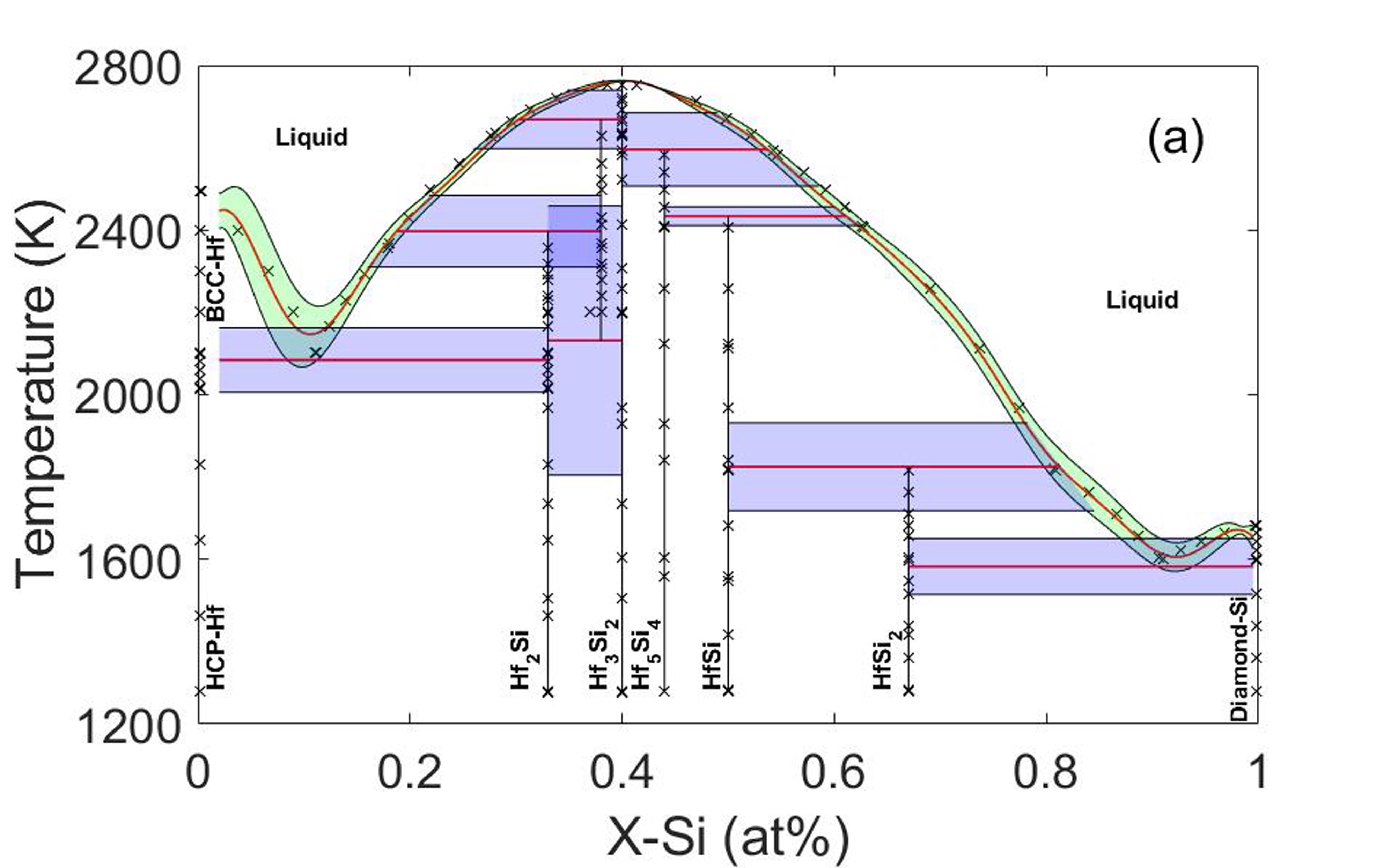}
  \label{Fig 8a}
\end{minipage}
\begin{minipage}{0.45\textwidth}
  \centering
  \includegraphics[width=1\linewidth]{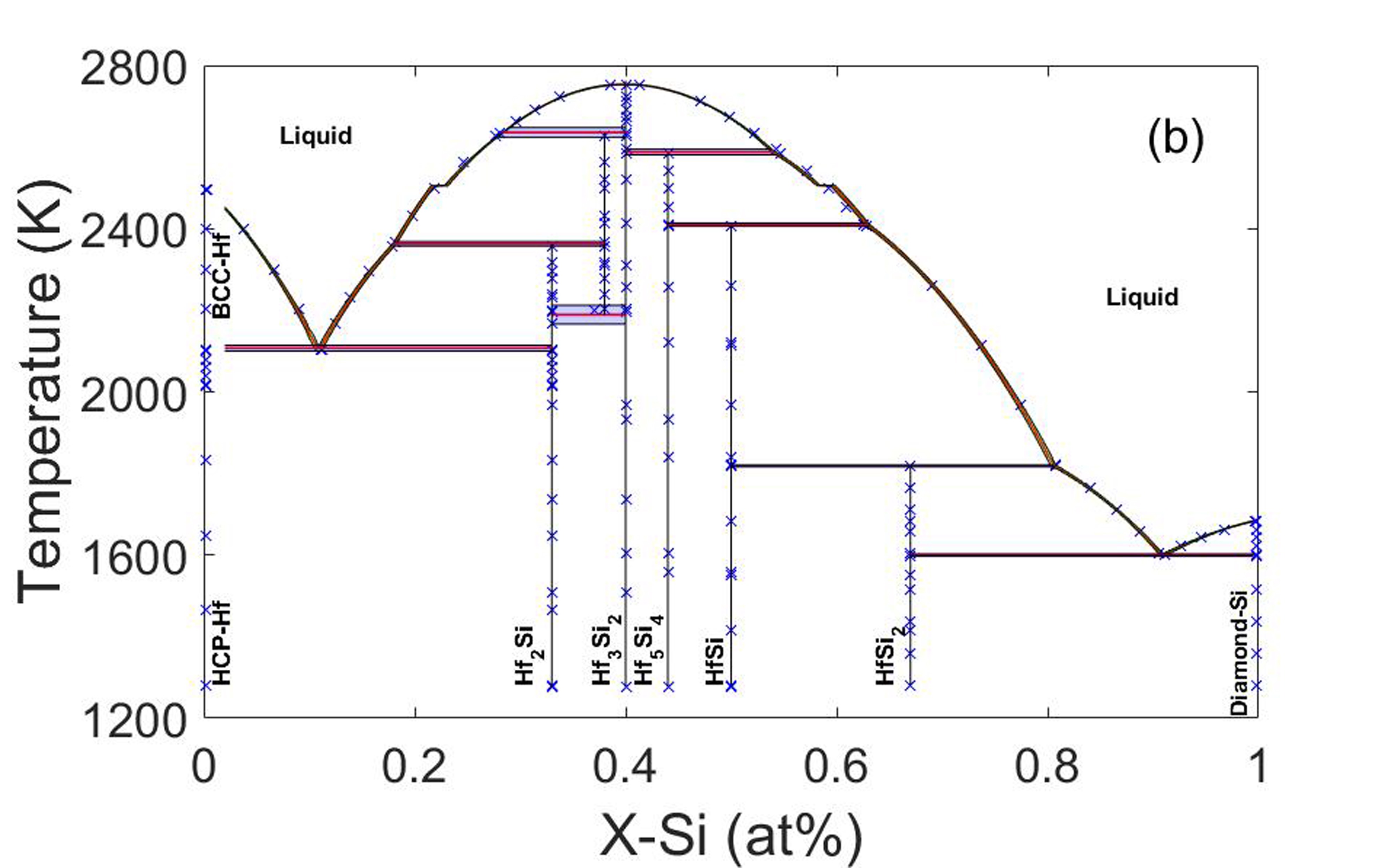}
\label{Fig 8b}
\end{minipage}
\caption{Error correlation-based model fusions of a) three models (1, 3, and 4) and b) all four models.\label{Fig 8}}
\end{figure*}

To perform this model fusion, the equivalent normal distributions at different compositions/regions are required to be calculated for each of the four models. Then, three and four models are fused based on the CMF approach, whose results are shown in Figure~\ref{Fig 8}. It should be noted that models 1, 3, and 4 with low precision and high uncertainties are first chosen for model fusion to examine whether the resulting fused model can be closer to the data and reduce the uncertainties. Figure~\ref{Fig 8}a shows that the approach can provide a phase diagram in much better agreement with data and less uncertainties compared to phase diagrams obtained from each one of the applied models individually. This result implies that random CALPHAD models can be fused together to find a reasonable estimation for phase diagram instead of trial-and-error to find the best predicting model. In addition, it is obvious that better predictions can be achieved as shown in Figure~\ref{Fig 8}b if model 2 (the best model) is also involved in the model fusion. However, it is very hard to compare the uncertainty of the fused model and model 2 based on the resulting phase diagrams. For this reason, the information (Shannon) entropies are calculated to quantify the model uncertainties. 


Shannon entropy can be utilized as a measure of uncertainty or missing information, which is determined as a weighted average (expected value) of information content gained from all the possible outcomes of an event. The information content obtained from an event outcome $i$ is defined as the negative logarithm of its probability, $-log(p_i)$. In this regard, it should be noted that event outcomes with lower probabilities convey more information since their occurrence are more surprising. In the case of $r$ discrete possible outcomes for an event, entropy can be defined as the following summation~\cite{wellmann_2012}:
\begin{equation}
\label{eq 19}
H = E[-log(p_i)] = - \sum_{i=1}^{r} p_i log(p_i).
\end{equation}
Here, the transformation temperature resulting from each individual or fused model at a specific composition/region is considered as an event. For each individual or fused model, the average of all the entropies associated with probability distributions over 95\% BCIs in different specified compositions/regions are introduced as the average entropy of that model:

\begin{equation}
\label{eq 20}
H_M = \frac{1}{m} \sum_{j=1}^{m} H_j,
\end{equation}
where $m$ is the total number of probability distributions in different specified compositions/regions, which is the same for all the models. The average entropies of the models are listed in Table~\ref{Table 3}. It can be observed that model 2 has lower average entropy/uncertainty than model 1, 3 or 4, as expected. 

The BMA fused model shows less average entropy/uncertainty compared to these three models. The lower uncertainty of the BMA fused model can be attributed to its concentrated distribution around the results of model 2, which gives more certainty to the total distributions over 95\% BCIs. 
However, the average model still has broader uncertainty bounds compared to each of the individual models, which can provide more confidence in robust design. In other words, broader uncertainties can give more assurance about the presence of specific microstructural phases corresponding to a phase diagram region of interest by more shrinking the safe design space of composition-temperature in that region. In addition, it seems that the BMA model can incorporate some additional information about the uncertainties in the phase diagram compared to model 2 (the best model).

The incorporation of uncertainty from the individual models through CMF can be more optimal than BMA due to the consideration of model error correlations, as the given thermodynamic models clearly are statistically correlated. According to Table~\ref{Table 3}, it is clear that the fused models can yield better predictions with less average entropies/uncertainties than the individual models used in each fusion case. 

\begin{table*}[htp]
\centering
\caption{Average entropy as a measure of uncertainty for each individual and fused model.}
\setlength{\arrayrulewidth}{1mm}
\setlength{\tabcolsep}{12pt}
\renewcommand{\arraystretch}{1.5}
\scalebox{0.62}
{\begin{tabular}{ c c c c c c c c }
\hline\hline
{} & Model 1 & Model 2 & Model 3 & Model 4 & Bayesian Average Model & Fused Model from 3 sources & Fused Model from 4 sources \\
\hline
$H_M$ & 1.8612 & 0.8442 & 1.9221 & 2.0122 & 1.6378 & 1.6254 & 0.7857 \\
\hline\hline
\end{tabular}}
\label{Table 3}
\end{table*}

\section{Summary and Conclusion}
\label{5}

Due to the importance of uncertainty quantification in CALPHAD, an MCMC sampling approach is utilized in this work for the probabilistic calibration of CALPHAD model parameters against the available data in the case of the Hf-Si binary system. Considering the vast high-dimensional parameter space in CALPHAD modeling, applying some prior information about the parameter values and ranges from Thermo-Calc optimization module is often required to achieve parameter convergence with a reasonable cost. However, choosing an appropriate CALPHAD model with a sufficient number of parameters is a challenging task. 

Therefore, a systematic approach is required to find an optimal model rather than a trial-and-error approach. For this purpose, Bayesian hypothesis testing (or Bayesian model selection) based on Bayes' factors is proposed in this work and applied in a case study for the Hf-Si system to show how the best model can be chosen from a list of the expert-proposed models. However, our work suggests to use information fusion approaches to smartly combine the given individual models into a fused model rather than just the application of the best model that may lose some useful information. BMA and an error correlation-based model fusion are used for our Hf-Si case study to show different beneficial purposes of these information fusion approaches.

The average model obtained from BMA shows larger 95\% confidence intervals compared to any one of the individual models, which can provide more confidence for robust design but is likely too conservative. On the other hand, the error correlation-based technique can provide closer results to data with less uncertainties than the individual models used for the fusion. The uncertainty reductions through this fusion approach are also verified through the comparison of the average entropies (as a measure of uncertainty) obtained for the individual and fused models. Therefore, random CALPHAD models can smartly be fused together to find reasonable predictions for phase diagrams with no need to go through the cumbersome task of identifying the best CALPHAD models. 

\section*{Acknowledgement}
Authors would like to thank the support of the National Science Foundation (NSF) [grant nos. CMMI-1534534,CMMI-1663130,DGE-1545403]. RA and DA also acknowledge the support of ARL through [grant No. W911NF-132-0018].

\bigskip
\textbf{References}
\bibliographystyle{elsarticle-num}
\bibliography{reference}

\end{multicols}
\end{document}